\documentclass[aps,pra,onecolumn,superscriptaddress]{revtex4-2}

\usepackage{url}
\usepackage[utf8]{inputenc}
\usepackage[T1]{fontenc}
\usepackage{graphicx} 

\usepackage{amsmath,amsthm,amssymb}
\usepackage{mathtools}

\usepackage{hyperref}
\usepackage[export]{adjustbox}

\usepackage{siunitx}
\DeclareSIUnit{\MHz}{\mega\hertz}
\DeclareSIUnit\year{yr}
\DeclareSIUnit{\rtHz}{\ensuremath{\mathrm{Hz}^{-1/2}}}

\usepackage{color}

\begin{document}

\title{Probing gravity at megahertz frequencies with levitated rotors}

\author{Matteo Fadel}
\email{matteofadel@pku.edu.cn}
\affiliation{Institute of Quantum Science and Technology $\&$ Frontiers Science Center for Nano-Optoelectronics, School of Physics, Peking University, Beijing 100871, China}
\affiliation{Department of Physics, ETH Z\"{u}rich, Z\"{u}rich 8093, Switzerland}

\author{Andrea Vinante}
\affiliation{Istituto di Fotonica e Nanotecnologie CNR-IFN, Trento 38123, Italy}
\affiliation{Fondazione Bruno Kessler (FBK), Trento 38123, Italy}

\begin{abstract}
The near-field gravitational interaction between masses is always thought of as instantaneous, an assumption experimentally tested only up to \qty{1.7}{\kilo\hertz}, where measurements yield a bounds at the \qty{e-5}{\second} level.
We propose a near-field experiment reaching up to megahertz frequencies, in which a levitated microrotor modulates a mass quadrupole and a second rotor measures the resulting gravitational tidal field as a torque. 
Because the detector rotor acts as a mechanical local oscillator, the megahertz signal is downconverted to a convenient band of a few hertz, and since at fixed geometry neither the torque nor the integration time depends on the rotation rate, higher frequencies are reached at no cost in sensitivity.
With parameters within reach of magnetically levitated NdFeB rotors, a signal at \qty{2}{\mega\hertz} yields a torque of \qty{2e-25}{\newton\metre}, detectable in under an hour at \qty{10}{\milli\kelvin}. 
Treating a possible departure from instantaneous gravitational coupling through a field relaxation time, the measured torque could constrain it at the \qty{e-8}{\second} level, improving on the existing bounds by more than three orders of magnitude. 
As a byproduct of our analysis, we derive the sensitivity of a rotor operated as a tunable high-frequency gravitational-wave detector, finding that a noise-equivalent strain around $10^{-22}$ Hz$^{-1/2}$ could be achievable in the megahertz band.
\end{abstract}

\maketitle
\tableofcontents

\clearpage
\newpage

\section{Introduction}

Gravity is possibly the least understood of the fundamental interactions, and it is by far the weakest, so that its study poses significant challenges. A compelling long-term goal is to probe its quantum nature in the laboratory, for instance through the gravity-induced entanglement of two masses prepared in spatial superposition states, but such experiments remain far from being feasible with current technology. Meanwhile, entire regions of the parameter space of the classical gravitational interaction have simply never been entered, and probing them requires no quantum state preparation at all.

One of these unexplored regions is the high-frequency one. Every laboratory measurement of the gravitational interaction between two masses has been, in effect, a quasi-static one, and the interpretation of all of them rests on the assumption that the near-field coupling is instantaneous. That assumption has never been tested at high frequency, although it plays an important role. The reason is largely technical, as the weakness of gravity at the micro and nanoscale makes any gravitational effect easily overwhelmed by electromagnetic interactions, and the schemes that achieve sufficient sensitivity are almost all slow.
Torsion balances operate at millihertz, and the Cavendish-type experiment on milligram masses of Ref.~\cite{westphal2021} modulates its source at \qty{12.7}{\milli\hertz}. The resonating-beam experiments of Refs.~\cite{brack2022,brack2023} work at \qty{42}{\hertz}, the levitated-oscillator schemes of Refs.~\cite{timberlake2021,fuchs2024,amaral2026} between \qty{27}{\hertz} and \qty{150}{\hertz}, and the milligram torsional pendulum of Ref.~\cite{Agafonova2016} operates at \qty{18}{\hertz}. Dynamic near-field experiments with rotating sources reach somewhat higher frequencies: a mass quadrupole spinning at \qty{460}{\hertz} was coupled to the cryogenic bar EXPLORER at twice that frequency \cite{astone1998}, following the earlier rotating-bar measurements of Refs.~\cite{hirakawa1980,kuroda1985}. Several experiments in the few kHz range are currently under development with the goal of measuring gravity with a detector oscillator in the mechanical ground state \cite{depellette2026,bonaldi2026}. At these frequencies, even ground-state cooling is challenging and nonclassical states have been partially realized only in levitated nanoparticles around \qty{100}{\kilo\hertz}. The highest frequency at which a gravitational interaction has ever been generated and detected in a laboratory remains the \qty{1660}{\hertz} of the acoustically stressed cylinder of Ref.~\cite{sinsky1968}, a result now nearly six decades old. Gravitational-wave observations are not a counterexample, since they probe radiative rather than near-field effects and in any case carry negligible signal above a kilohertz \cite{Ligo2015}. The entire range above this frequency is experimentally largely untested.

In this work, we propose the realization of a two-mass near-field gravitational experiment at unprecedented high frequency up to several megahertz. In our proposal, the source of gravity is a microscale rotor operated close to the maximum frequency allowed by the material \cite{schuck2018,jose2025}. For a rotor with a length of 50-100 micrometers, this can be up to \qty{5}{\mega\hertz}, leading to gravitational fields modulated at up to \qty{10}{\mega\hertz}. As detector we will consider a second rotor operated as a heterodyned torque detector. Its own rotation acts as a mechanical local oscillator, so that the megahertz signal is downconverted to a band of a few hertz. This is what distinguishes this proposal from every resonant scheme \cite{astone1998,goryachev2014}, in which the accessible frequency is pinned to a mechanical mode of the detector rather than chosen through an experimentally tunable parameter.

Extending the frequency range of gravitational experiments opens a new way to test the near-field instantaneity assumed above. A broad class of scenarios, in which gravity is emergent or thermodynamic in origin \cite{jacobson1995,verlinde2011} or in which an additional gravitational degree of freedom carries a dispersive self-interaction, predicts a departure from instantaneity at a timescale that the frameworks themselves do not fix \cite{diosi2013,diosi2014}. We treat that timescale as a free phenomenological parameter to be bounded from experimental measurements.

Two features make our scheme well suited to this purpose. First, for a rotor detector the predicted signal and the integration time required to reach a given signal-to-noise are both independent of the rotation rate at fixed geometry, so that higher frequencies are reached without penalty in sensitivity. Second, the rotation rate of a levitated rotor is continuously tunable up to the structural limit, so that a single apparatus can map the gravitational coupling across several orders of magnitude in frequency. Any departure from Newtonian behaviour would then appear as a deviation from a flat response in the measured amplitude, and as a growing offset in the measured phase, the latter being referenced to the independently monitored angular positions of the two rotors. We emphasize that the scheme is entirely classical, which is precisely why it can be assessed with existing numbers and attempted with existing hardware. Preparing the rotors in nonclassical states, which would bring the scheme within reach of the quantum questions raised above, is a natural extension that we leave to future work.

As a further outcome of our analysis, we derive the sensitivity of a rotor as a torque-based gravitational-wave detector. In fact, the field gradient generated by a rotor source along the rotation axis has the same form as that of a circularly polarized gravitational wave propagating along the axis. We find that a strain noise $\sqrt{S_h}<10^{-22}$~Hz$^{-1/2}$ could be achievable in the megahertz-frequency regime near the rotor structural limit, competitive with or better than other approaches. Despite being narrowband, such a detector would feature two unique properties: it would be directly sensitive to circularly polarized signals, and its central frequency would be fully tunable.

\section{Gravity sources}
\label{sec:sources}

A stationary mass distribution produces a gravitational signal at zero frequency, where the output of any mechanical detector is dominated by drifts and by fluctuations whose power spectral density rises as $1/f$, including thermal expansion of the experiment, seismic tilts, and slow relaxation of clamping stresses. 
Such contributions are correlated over subsequent measurements, so the signal-to-noise ratio does not improve as the square root of the integration time and the noise floor cannot be reduced by averaging alone. Modulating the source mass brings the signal to a band in which the noise is white, permitting phase-coherent demodulation against an independently measured reference \cite{westphal2021}. 
Among the available modulation schemes, continuous rotation of a rigid body has a number of distinctive features. 
First, the amplitude of the source moment is a property of the mass distribution and is independent of the modulation frequency, whereas a reciprocating (i.e. moving back-and-forth) mass requires a drive force growing quadratically with frequency at fixed displacement. 
Second, rotation is unconstrained by any elastic restoring force, so no stored energy must be exchanged each cycle. The phase and frequency of the emitted signal are determined by the angular position of the rotor, which is measurable to high accuracy and provides the reference required for coherent detection over integration times limited only by the stability of the rotation itself. 
Rotating masses are, for these reasons, a well established tool in precision gravity experiments, such as in subcentimeter tests of the inverse-square law \cite{kapner2007,adelberger2003} and in the Newtonian calibrators that set the absolute displacement scale of gravitational-wave interferometers \cite{estevez2018,inoue2018,ross2021}.

The multipole order accessible to such a source is fixed by conservation laws. Mass conservation renders the monopole time independent, and momentum conservation renders the dipole time independent in the center-of-mass frame, which is the frame in which a freely levitated body rotates. 
The quadrupole is therefore the lowest multipole that an unsupported rotor can modulate, and for a body spun about a principal axis it oscillates at twice the rotation frequency. 
The attainable frequency is limited by the tensile strength of the rotor material through the specific strength, since the maximum tip speed of a rotating solid is set by the square root of the ratio of breaking stress to density and does not exceed several hundred meters per second for available materials \cite{schuck2018,jose2025}. 
The maximum angular velocity consequently scales inversely with the linear dimension of the rotor, and frequencies in the megahertz to gigahertz range require source dimensions in the range of tens to hundreds of micrometers. 
Rotation rates above one gigahertz have been demonstrated in optically levitated nanoparticles \cite{reimann2018,ahn2018,jin2021}, and megahertz rates in magnetically levitated micrometre-scale rotors \cite{schuck2018,jose2025}, exceeding by several orders of magnitude what is accessible to resonant or reciprocating masses. Because the near-field gravitational signal depends on the source dimension and the source-detector separation only through their ratio, miniaturization incurs no penalty in amplitude provided the detector can be positioned proportionately closer. In the following we quantify these statements, beginning with the multipole expansion of the Newtonian potential and concluding with the field-gradient tensor at the position of a nearby detector. We close the section by treating, for comparison, a reciprocating source mass.

\subsection{Newtonian potential of a mass quadrupole}
\label{sec:nearfield_quadrupole}

Consider a mass distribution $\rho(\mathbf{x},t)$ confined within a region of characteristic size $l$ centered on the origin. 
At a distance $r \equiv |\mathbf{r}| \gg l$, the resulting Newtonian potential is
\begin{equation}
\Phi(\mathbf{r},t) = -G \int d^3x \, \frac{\rho(\mathbf{x},t)}{|\mathbf{r}-\mathbf{x}|} \;.
\label{eq:Phi_def}
\end{equation}
Let $\mathbf{n} \equiv \mathbf{r}/r$ denote the unit vector pointing toward the field point $\mathbf{r}$. 
Taylor expanding the Newtonian kernel about $\mathbf{x} = 0$, using the shorthand notation $\partial_i \equiv \partial/\partial r_i$ and implicit summation over repeated indices throughout, we have
\begin{equation}
\frac{1}{|\mathbf{r}-\mathbf{x}|} = \frac{1}{r} - x_i \partial_i\left(\frac{1}{r}\right) + \frac{1}{2} x_i x_j \partial_i \partial_j\left(\frac{1}{r}\right) + \mathcal{O}\left(\frac{l^3}{r^4}\right) \;.
\label{eq:greens_expand}
\end{equation}
From $\partial_i\left(1/r\right) = - n_i / r^2$, $\partial_i\partial_j\left(1/r\right) = (3n_i n_j - \delta_{ij})/r^3$, we then obtain
\begin{equation}
\Phi(\mathbf{r},t) = -G \int d^3x \, \rho(\mathbf{x},t) \left[ \frac{1}{r} + \frac{x_i n_i}{r^2} + \frac{1}{2}\frac{x_i x_j\,(3n_i n_j - \delta_{ij})}{r^3} + \mathcal{O}\left(\frac{l^3}{r^4}\right) \right].
\label{eq:Phi_expand_raw}
\end{equation}
The first two moments of the source are the total mass and the mass dipole,
\begin{equation}
M \equiv \int d^3x \, \rho(\mathbf{x},t), \qquad D_i(t) \equiv \int d^3x \, \rho(\mathbf{x},t)\, x_i .
\label{eq:M_D_def}
\end{equation}
Mass conservation makes $M$ time independent, and for an isolated source $\ddot{D}_i = 0$. Placing the origin at the center of mass in the frame where the total momentum vanishes therefore enforces $D_i(t) = 0$ at all times, so that the leading time-dependent contribution to the Newtonian potential $\Phi(\mathbf{r},t)$ arises at quadrupole order.

Because $\mathbf{n}$ is a unit vector, only the symmetric trace-free part of the second moment survives the contraction with $n_i n_j$. Explicitly, $x_i x_j\,(3 n_i n_j - \delta_{ij}) = \left(3 x_i x_j - |\mathbf{x}|^2 \delta_{ij}\right) n_i n_j$. 
This motivates the mass quadrupole tensor
\begin{equation}
Q_{ij}(t) \equiv \int d^3x \, \rho(\mathbf{x},t) \left(3 x_i x_j - |\mathbf{x}|^2 \delta_{ij}\right),
\label{eq:Qij_def}
\end{equation}
which is symmetric and trace free, i.e. $Q_{ii}(t) = 0$, by construction. Conventions differ by the placement of the factor of $3$, here $Q_{ij}(t)$ is three times the trace-free second moment. 
The exterior potential up to and including quadrupole order is then
\begin{equation}
\Phi(\mathbf{r},t) = -\frac{GM}{r} - \frac{G}{2r^3}\, Q_{ij}(t) \, n_i n_j + \mathcal{O}\left(\frac{GMl^3}{r^4}\right).
\label{eq:Phi_quadrupole}
\end{equation}
The first neglected term is the octupole contribution, smaller than the quadrupole term by a further power of $l/r$.

The gravitational field and the tidal (field-gradient) tensor follow from
\begin{equation}
g_i(\mathbf{r},t) = -\partial_i \Phi(\mathbf{r},t), \qquad G_{ij}(\mathbf{r},t) = \partial_i g_j(\mathbf{r},t) = -\partial_i \partial_j \Phi(\mathbf{r},t),
\label{eq:g_Gij_defs}
\end{equation}
with $G_{ij} = G_{ji}$ and, in vacuum, $G_{ii} = 0$ since $\nabla^2 \Phi = 0$. 
The monopole alone gives the familiar static tide
\begin{equation}
G_{ij}^{(\rm{m})}(\mathbf{r}) = \frac{GM}{r^3}\left(3 n_i n_j - \delta_{ij}\right),
\label{eq:Gij_monopole}
\end{equation}
so any time dependence of the local tidal field at point $\mathbf{r}$ is carried entirely by $Q_{ij}(t)$. 
This is the term a rotor modulates, and the focus of the following sections.

\begin{figure}
    \centering
    \includegraphics[width=0.7\linewidth]{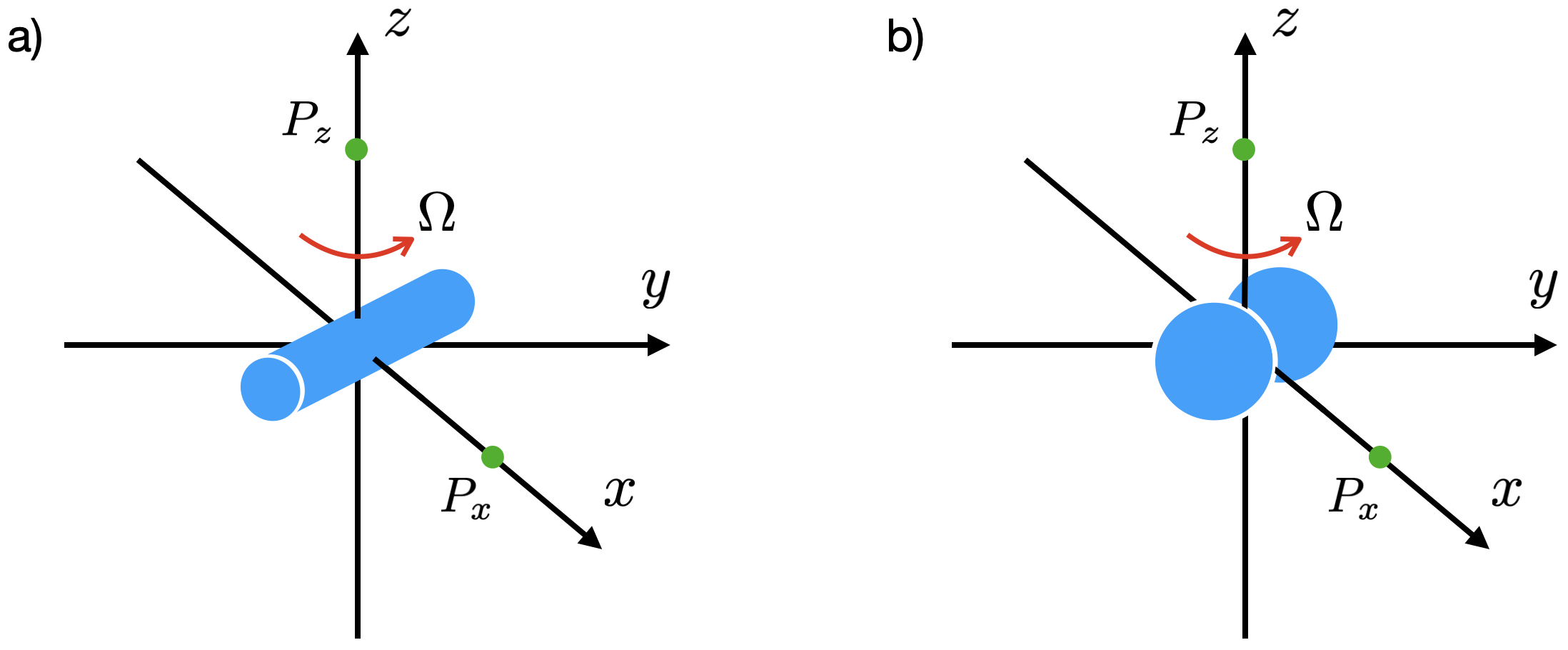}
    \caption{\textbf{Illustration of spinning rotors.} We consider the gravitational field generated by \textbf{a)} a cylindrical rod of mass density $\rho$, length $l$, and radius $a$, and by \textbf{b)} a dumbbell made of two touching spheres of radius $R$ and mass density $\rho$, that are spinning at angular frequency $\Omega$ around the vertical, $z-$axis. We compute the gravitational field experienced at points $P_x=(r,0,0)$ and $P_z=(0,0,r)$ situated at distance $r$ from the origin along the $x-$ and $z-$axis, respectively.}
    \label{figRotor}
\end{figure}

\subsection{Near-field gradient from a spinning rotor}
\label{sec:cylinder}

We now specialize the multipole expansion of Sec.~\ref{sec:nearfield_quadrupole} to a laboratory source of realistic realization, namely a rigid body spun about a fixed axis, see Fig.~\ref{figRotor}. Throughout, note that we are concerned with the quasi-static Newtonian near field of the source rather than with the far weaker radiative field (i.e. gravitational wave) it emits.

\subsubsection{Quadrupole of a rigid rotor}

Let the rotor spin about the $\hat{\mathbf{z}}$ axis at angular velocity $\Omega$, and let $I_1$ and $I_2$ be the principal moments of inertia about the two body axes orthogonal to the spin axis, with $\hat{\mathbf{u}}(t)=(\cos(\Omega t) ,\sin(\Omega t),0)$ the instantaneous lab-frame direction of the first of them. Writing the body-frame second-moment tensor in its principal basis and rotating it into the lab frame, the constant part of $Q_{ij}(t)$ decouples and the time-dependent part takes the form
\begin{equation}
Q_{ij}(t) = Q \begin{pmatrix} \cos (2\Omega t) & \sin (2\Omega t) & 0 \\ \sin (2\Omega t) & -\cos (2\Omega t) & 0 \\ 0 & 0 & 0 \end{pmatrix}, \qquad Q = \frac{3}{2}\left(I_2 - I_1\right) \equiv \frac{3}{2}\,\Delta I .
\label{eq:Qrotor}
\end{equation}
Only the transverse asymmetry $\Delta I$ enters, the signal oscillates at twice the rotation frequency, and a rotor that is symmetric about the spin axis produces no time-dependent near field. 
For an axially symmetric body spun about a transverse axis, which is the configuration of interest here, Eq.~\eqref{eq:Qrotor} reduces to $Q=\tfrac{3}{2}(I_\perp - I_\parallel)$ with $I_\parallel$ the moment of inertia about the symmetry axis and $I_\perp$ that about a transverse axis through the center of mass. 
This single relation conveniently replaces the explicit volume integrals in every case considered below.

\subsubsection{Cylinder and dumbbell}

Consider first a homogeneous cylinder of density $\rho$, radius $a$ and length $l$, with its symmetry axis lying in the $xy$ plane and rotating in that plane at frequency $\Omega$, as shown in Fig.~\ref{figRotor}a. 
Using $I_\parallel = Ma^2/2$ and $I_\perp = M(l^2/12 + a^2/4)$ with $M=\pi\rho a^2 l$, Eq.~\eqref{eq:Qrotor} gives
\begin{equation}
Q_{\rm cyl} = \frac{\pi}{8}\,\rho\,\alpha^2\left(1-3\alpha^2\right) l^5, \qquad \alpha \equiv \dfrac{a}{l} .
\label{eq:Qcyl}
\end{equation}
We now optimize the aspect ratio $\alpha$ at fixed length rather than at fixed mass, because for an elongated rotor the breaking stress, and hence the maximum attainable spin frequency, is controlled mostly by its length $l$ (see Sec.~\ref{sec:limits}).
At fixed $l$ the function $\alpha^2(1-3\alpha^2)$ peaks at $\alpha=1/\sqrt{6}$, meaning at $l \approx 2.4 \,a$, giving the optimized value
\begin{equation}
Q_{\rm cyl}^\ast = \frac{\pi}{96}\,\rho\, l^5 .
\label{eq:Qmax}
\end{equation}

Note that Eq.~\eqref{eq:Qcyl} is positive for $\alpha<1/\sqrt{3}$, which is the condition under which the rotor is prolate (cigar-like, in the inertial sense) about its symmetry axis.
Since $Q=\tfrac{3}{2}(I_\perp-I_\parallel)=\tfrac{M}{8}\left(l^{2}-3a^{2}\right)$, the quadrupole vanishes precisely where the body becomes an inertially spherical top, which is also where the spin axis ceases to be an axis of maximum inertia.

The second case we consider is a dumbbell made of two identical touching spheres of radius $R$ and total length $l=4R$, as shown in Fig.~\ref{figRotor}b. 
It has $I_\perp - I_\parallel = 2mR^{2}$ for the pair, with $m=\tfrac{4}{3}\pi\rho R^{3}$ the mass of one sphere, so that its quadrupole has exactly the form Eq.~\eqref{eq:Qrotor} with
\begin{equation}
Q_{\rm db} = 3mR^2 = \frac{\pi}{256}\,\rho\, l^5 .
\label{eq:Qdb}
\end{equation}
At fixed density and fixed total length $l$ the optimal cylinder is therefore superior by 
\begin{equation}
    \dfrac{Q_{\rm cyl}^\ast}{Q_{\rm db}} = \dfrac{8}{3} \;,
\end{equation}
and it is also far more robust against disintegration, so it can be spun to higher frequencies. 
This comparison is made at fixed $\rho$ and $l$ rather than at fixed mass, precisely because the maximum attainable frequency is set by the length. Ranking the two geometries per unit mass would reverse the conclusion. Since $M_{\rm cyl}^\ast=\pi\rho l^3/6$ and $M_{\rm db}=\pi\rho l^3/24$ give $M_{\rm cyl}^\ast=4M_{\rm db}$, one has
\begin{equation}
\frac{Q_{\rm cyl}^\ast}{M_{\rm cyl}^\ast} = \frac{l^2}{16}, \qquad \frac{Q_{\rm db}}{M_{\rm db}} = \frac{3 l^2}{32},
\label{eq:QoverM}
\end{equation}
corresponding to a factor $3/2$ in favor of the dumbbell. Which figure of merit is the relevant one depends on whether the experiment is limited by the available mass or by the attainable spin rate. 
For the levitated microrotors considered here we take the latter as main limitation.

\subsubsection{Field gradient at an arbitrary field point}

\begin{figure}
    \centering
    \includegraphics[width=0.4\linewidth, valign=c]{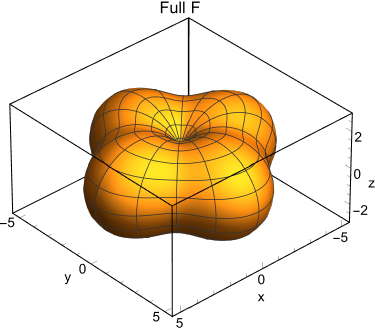}
    \includegraphics[width=0.45\linewidth, valign=c]{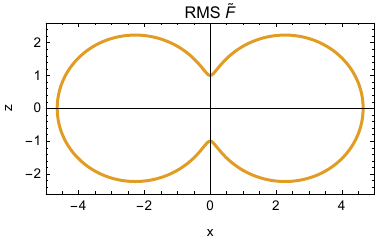}
    \caption{\textbf{Angular pattern of the near-field tidal amplitude for a spinning rotor.} Normalized amplitude of the field-gradient tensor at fixed distance from the rotor. Both patterns are normalized to unity on the spin axis, so the plotted radius is the enhancement relative to the axial value. Left: the instantaneous pattern $F(\theta,\varphi,t)$ of Eq.~\eqref{eq:Ginvariant}, shown at fixed phase. The four lobes reflect the $m=2$ structure of the rotating quadrupole and are rigidly attached to the source, sweeping around the spin axis at $\Omega$. Their period of $\pi$ in azimuth is the origin of the signal at twice the rotation frequency. The pattern reaches $F=\tfrac{1}{2}\sqrt{109}\simeq 5.22$ along the lobe axes and $F=4$ between them, so the azimuthal contrast is a factor $1.30$ and there are no nodes. Right: the meridional section of the cycle-averaged pattern $\widetilde{F}(\theta)$ of Eq.~\eqref{eq:pattern}, obtained by averaging $F^{2}$ over one rotation period. The azimuthal dependence is thereby removed and the pattern becomes axially symmetric, so the full three-dimensional surface is the figure of revolution generated by the curve shown. The amplitude grows monotonically from $\widetilde{F}=1$ on the spin axis to $\widetilde{F}=\sqrt{173/8}\simeq 4.65$ in the equatorial plane, which is thus the preferred location for a detector.}
    \label{figPattern}
\end{figure}

Inserting the time-dependent quadrupole term of Eq.~\eqref{eq:Phi_quadrupole} into the definition Eq.~\eqref{eq:g_Gij_defs} gives
\begin{equation}
G_{ij}(\mathbf{r},t) = \frac{G}{2}\, Q_{kl}(t)\, \partial_i \partial_j \left(\frac{n_k n_l}{r^3}\right) = \frac{G}{2}\, Q_{kl}(t)\, \partial_i \partial_j \left(\frac{x_k x_l}{r^5}\right).
\label{eq:Gij_full}
\end{equation}
Here we do not need to consider the term $-GM/r$ in the potential Eq.~\eqref{eq:Phi_quadrupole}, as it produces only a static offset in the tidal tensor, i.e.\ a constant force gradient on the test mass.
Carrying out the differentiation and contracting with a symmetric trace-free $Q_{kl}(t)$, all four nonvanishing components of Eq.~\eqref{eq:Qrotor} contributing, yields the closed form
\begin{equation}
G_{ij}(\mathbf{r},t) = \frac{G}{r^5}\left[ Q_{ij}(t) - 5\left(n_i Q_{jk}(t) n_k + n_j Q_{ik}(t) n_k\right) - \frac{5}{2}\,\delta_{ij}\, Q_{kl}(t) n_k n_l + \frac{35}{2}\, n_i n_j\, Q_{kl}(t) n_k n_l \right],
\label{eq:Gij_general}
\end{equation}
which is manifestly symmetric and, using $Q_{ii}(t)=0$, trace free. Equation~\eqref{eq:Gij_general} holds for any field direction and makes two cases of interest immediate: one along the spin axis and another along a fixed equatorial direction. These are the points $P_z$ and $P_x$, respectively, in Fig.~\ref{figRotor}, which we take at a distance $r$ from the rotor center.

For the axial case $\mathbf{r}=(0,0,r)\equiv P_z$ one has $Q_{ik}(t)n_k=0$ and $Q_{kl}(t)n_kn_l=0$, so only the first term survives,
\begin{equation}
G_{ij}(P_z,t) = \frac{GQ}{r^5} \begin{pmatrix} \cos (2\Omega t)  & \sin (2\Omega t)  & 0 \\ \sin (2\Omega t)  & -\cos (2\Omega t)  & 0 \\ 0 & 0 & 0 \end{pmatrix}_{\! ij} .
\label{eq:Gijz}
\end{equation}
For the equatorial case $\mathbf{r}=(r,0,0)$ all four terms contribute and
\begin{equation}
G_{ij}(P_x,t) = \frac{GQ}{r^5} \begin{pmatrix} 6\cos (2\Omega t)  & -4\sin (2\Omega t)  & 0 \\ -4\sin (2\Omega t)  & -\frac{7}{2}\cos (2\Omega t)  & 0 \\ 0 & 0 & -\frac{5}{2}\cos (2\Omega t)  \end{pmatrix}_{\! ij} .
\label{eq:Gijx}
\end{equation}
Both tensors are trace free, as required in vacuum, and the equatorial amplitudes are substantially larger, as expected.

The full angular dependence can be given in analytical form. For a general direction $\mathbf{n}=(\sin\theta\cos\varphi,\sin\theta\sin\varphi,\cos\theta)$ the two scalars entering Eq.~\eqref{eq:Gij_general} are
\begin{equation}
Q_{ik}(t)n_{k}Q_{il}(t)n_{l} = Q^{2}\sin^{2}\theta, \qquad Q_{kl}(t)n_{k}n_{l} = Q\sin^{2}\theta\,\cos[2 (\Omega t - \varphi)] \;,
\label{eq:contractions}
\end{equation}
so that the azimuth and the time enter only through the single phase $(\Omega t - \varphi)$. This illustrates the fact that a displacement of the field point in azimuth is equivalent to a translation in time. The squared Frobenius norm of the tidal tensor then follows in closed form,
\begin{equation}
G_{ij}G_{ij} = \frac{2G^{2}Q^{2}}{r^{10}}\, F^{2}(\theta,\varphi,t), \qquad\qquad F(\theta,\varphi,t) \equiv \sqrt{ 1 + 15\sin^{2}\theta + \frac{45}{4}\sin^{4}\theta\,\cos^{2} [2(\Omega t - \varphi)] } ,
\label{eq:Ginvariant}
\end{equation}
where $F$ is normalized to unity on the spin axis, see Fig.~\ref{figPattern}. The emission pattern $F(\theta,\varphi,t)$ is a function rigidly attached to the source and thus rotating with it at angular velocity $\Omega$. Its period of $\pi$ in $(\Omega t - \varphi)$ is the origin of the factor of two between the rotation frequency $\Omega$ and the signal frequency $\omega=2\Omega$. In the equatorial plane it ranges from $F=4$ in the directions between the lobes to $F=\sqrt{109}/2\simeq 5.22$ along the lobe axes, so the azimuthal contrast is only a factor $1.30$ and the pattern has no nodes. The axial value $F=1$ is a minimum rather than a zero, the field there being nonvanishing and constant in time.

Averaging $F^{2}(\theta,\varphi,t)$ over time removes the phase dependence and leaves an axially symmetric pattern
\begin{equation}
\widetilde{F}(\theta) \equiv \left\langle F^{2}(\theta,\varphi,t)\right\rangle_t^{1/2} = \sqrt{ 1 + 15\sin^{2}\theta + \frac{45}{8}\sin^{4}\theta },
\label{eq:pattern}
\end{equation}
with $\widetilde{F}(0)=1$ and $\widetilde{F}(\pi/2)=\sqrt{173/8}\simeq 4.65$, see Fig.~\ref{figPattern}. 
The root-mean-square tidal amplitude is therefore larger in the equatorial plane than on the spin axis by a factor $4.65$, to be compared with the factor $6$ obtained from the single largest component of Eq.~\eqref{eq:Gijx}. Both $F$ and $\widetilde{F}$ characterize the source rather than any particular probe. A detector couples to a single projection of $G_{ij}$ selected by its mode geometry, so its response depends on its orientation as well as on its position.

Note that averaging $F^{2}$ rather than $F$ is the appropriate operation, since every component of $G_{ij}$ oscillates about zero at a fixed field point and the linear cycle average vanishes identically. For the monochromatic signal considered here the root mean square is proportional to the amplitude, and it is the quantity that determines the signal-to-noise ratio accumulated by a narrowband detector integrating coherently. Because the phase enters only through $(\Omega t - \varphi)$, the average over a rotation period at fixed azimuth coincides with the average over azimuth at fixed time, which is why $\widetilde{F}$ depends on the polar angle $\theta$ only.

\subsubsection{Equivalent gravitational-wave strain}
\label{sec:heq}

It is useful to express the Newtonian field gradient of the rotor as the strain of the gravitational wave that would produce the same tidal effect, since detector sensitivities in this frequency range are conventionally quoted as strain. For a monochromatic plane wave in the transverse-traceless gauge the geodesic deviation of two free test masses is governed by $G^{\rm GW}_{ij}=\tfrac{1}{2}\ddot{h}_{ij}$. 
A circularly polarized wave $h_{ij}(t) = h\left(\begin{smallmatrix} \cos(\omega t)  & \sin(\omega t) & 0 \\ \sin(\omega t) & -\cos(\omega t)  & 0 \\ 0 & 0 & 0 \end{smallmatrix}\right)$, of amplitude $h$ and angular frequency $\omega$, carries a tidal tensor of squared Frobenius norm
\begin{equation}
    G^{\rm GW}_{ij}G^{\rm GW}_{ij}=\dfrac{1}{2}\omega^{4}h^{2} \;.
\label{eq:GGgw}
\end{equation}
Equating Eq.~\eqref{eq:GGgw} to that of the rotor field Eq.~\eqref{eq:Ginvariant}, time averaged as in Eq.~\eqref{eq:pattern} and taking into account that spinning at $\Omega$ produces a signal at $\omega=2\Omega$, gives the equivalent strain as a function of direction
\begin{equation}
h_{\rm eq}(\theta) = \frac{2\,G Q}{\omega^{2}r^{5}}\,\widetilde{F}(\theta) = \frac{3\,G\,\Delta I}{\omega^{2}r^{5}}\,\widetilde{F}(\theta).
\label{eq:heq_general}
\end{equation}
The second equality follows from $Q=\tfrac{3}{2}\Delta I$ and it expresses the signal entirely through the transverse inertia asymmetry of the rotor $\Delta I$. The angular factor is always the pattern Eq.~\eqref{eq:pattern} plotted in Fig.~\ref{figPattern}.
Note that taking a linearly rather than circularly polarized reference wave would change the right-hand side of Eq.~\eqref{eq:GGgw} into $\omega^{4}h^{2}/4$, thus increasing Eq.~\eqref{eq:heq_general} by a factor $\sqrt{2}$.

On the spin axis the identification Eq.~\eqref{eq:heq_general} is actually exact. The tensor Eq.~\eqref{eq:Gijz} is traceless and satisfies $n_{i}G_{ij}=0$ for $\mathbf{n}=\hat{\mathbf{z}}$, so it is transverse-traceless with respect to the radial direction, and its two independent components are in quadrature. It is therefore indistinguishable from a circularly polarized wave propagating along $z$ at $\omega=2\Omega$, and any detector would respond to the two identically. Away from the axis no such identification exists. The tensor Eq.~\eqref{eq:Gijx} is traceless but not transverse, since $n_{i}G_{ij}\neq 0$ for $\mathbf{n}=\hat{\mathbf{x}}$, and its largest component $G_{xx}=6GQ/r^{5}$ is in fact the longitudinal one. 
This is a generic property of a near field, since only radiative solutions are transverse in every direction.
Equation~\eqref{eq:heq_general} is consequently a statement about ``effective strain magnitude'' and not about tensor structure, so two fields with equal $h_{\rm eq}$ need not drive a given detector equally. 
For example, a detector aligned with the dominant axis of Eq.~\eqref{eq:Gijx} sees $12GQ/\omega^{2}r^{5}$, exceeding the equatorial value of Eq.~\eqref{eq:heq_general} by $29\%$, the two differing by how much of the source's anisotropy the chosen projection retains.

A further consideration comes from the fact that a gravitational wave results in a spatially uniform tidal tensor across a detector of size $d$, whereas the rotor field varies over the detector by a relative amount of order $d/r$. The strain Eq.~\eqref{eq:heq_general} is therefore the leading term of an expansion in $d/r$, and for a detector comparable in size to its distance from the source the higher gradients must be retained. This is the same restriction that limits the multipole expansion of Sec.~\ref{sec:nearfield_quadrupole} to $r\gg l$, but here applied to the detector rather than to the source.

Using the optimal-cylinder moment Eq.~\eqref{eq:Qmax}, the strain in the equatorial plane evaluates to
\begin{equation}
h_{\rm eq}^{\rm cyl}(\pi/2) = \frac{\pi\,\widetilde{F}(\pi/2)}{48}\,\frac{G\rho}{\omega^{2}}\left(\frac{l}{r}\right)^{5} \simeq 0.304\,\frac{G\rho}{\omega^{2}}\left(\frac{l}{r}\right)^{5},
\label{eq:h_opt}
\end{equation}
so that the rotor dimensions enter only through the ratio $l/r$. At fixed signal frequency the strain is therefore invariant under a rescaling of the entire apparatus (source size and source-probe separation), even though such a rescaling changes the mass of the rotor as $l^{3}$. The relevant material property is the density rather than the mass. 
This invariance holds at fixed $\omega$ alone, since the maximum rotation frequency scales as $1/l$, meaning that for a rotor operated at its structural limit the absolute size reappears, as we are going to show later in Eq.~\eqref{eq:h_material}. 
Taking $\rho=\qty{7430}{\kilo\gram\per\cubic\metre}$, the density of the magnetically levitated NdFeB rotor of Ref.~\cite{jose2025}, Eq.~\eqref{eq:h_opt} evaluates to
\begin{equation}
h_{\rm eq}^{\rm cyl, NdFeB} = \num{3.8e-21}\left(\frac{\qty{1}{\MHz}}{f}\right)^{2}\left(\frac{l}{r}\right)^{5},
\label{eq:h_numeric}
\end{equation}
with $f=\omega/2\pi=2f_{\rm rot}$ the signal frequency and $f_{\rm rot}=\Omega/2\pi$ the rotation frequency. 

Two reference values bracket the range of interest, both evaluated at $r=2l$ so that $(l/r)^{5}=1/32$. A rotor at $f_{\rm rot}=\qty{3}{\MHz}$, which by Eq.~\eqref{eq:steel_limit} requires $l\lesssim\qty{100}{\micro\metre}$, gives $h_{\rm eq}^{\rm cyl, NdFeB}\simeq\num{3.3e-24}$ at $f_1=\qty{6}{\MHz}$. A rotor at $f_{\rm rot}=\qty{1}{\giga\hertz}$, for which the scaling $\Omega_{\max}\propto 1/l$ requires $l\lesssim\qty{300}{\nano\metre}$ (attainable with optically levitated nanorods), gives $h_{\rm eq}^{\rm cyl, NdFeB}\simeq\num{3.0e-29}$ at $f_2=\qty{2}{\giga\hertz}$. 
The reduction by five orders of magnitude between the two is exactly $(f_{2}/f_{1})^{2}$, since at fixed density and fixed $l/r$ the tidal field itself is unchanged, and the entire reduction originates from the $\omega^{-2}$ scaling of the conversion Eq.~\eqref{eq:heq_general} rather than any weakening of the underlying Newtonian field.

\subsubsection{Scaling, validity and material limits}
\label{sec:limits}

Equation~\eqref{eq:h_numeric} exhibits two features that control the feasibility of our proposed experiment. The first is the $h_{\rm eq} \sim \omega^{-2}$ scaling which, as mentioned in the previous section, is just a property of the gradient-to-strain conversion Eq.~\eqref{eq:heq_general} and not a suppression of the Newtonian signal.
In fact, the gravity-gradient amplitude produced by a rigid rotor, $G_{ij}\sim GQ/r^5$, does not itself decrease with frequency at fixed geometry. What decreases is the equivalent strain $h_{\rm eq}$, because a wave of fixed $h$ produces a tidal gradient growing as $\omega^2 h$, so that matching a given Newtonian gradient necessarily gives $h_{\rm eq}\propto G_{ij}/\omega^2$.  

The second feature is the fifth-power suppression with distance, which requires a detector as close to the source as possible. This conflicts with the multipole expansion itself, which requires $r\gg l$ up to factors of order unity. For $r$ only a few times $l$, and especially for $r\sim l$, the quadrupole approximation is no longer under control and the exact Newtonian field of the finite rotor must be used. 
For the symmetric geometries considered here the odd multipoles vanish, so the leading correction is hexadecapolar, of relative order $(l/2r)^{2}$ in the rotor half-length. We take the residual error of the quadrupole treatment there to be of order $\sim10\%$, as it can be found from a numerical evaluation of the exact finite-body field.

A notable feature of Eq.~\eqref{eq:h_numeric} is that only the ratio $l/r$ enters, not the absolute scale. The absolute scale is nevertheless fixed indirectly, because the highest attainable frequency is limited by mechanical disintegration. 
For a uniform rod rotating about its center the axial tensile stress is maximal at the center and equals $\sigma_{\max}=F_{\rm ct}/A=\rho\Omega^2 l^2/8$, where $F_{\rm ct}=(M/2)\Omega^2(l/4)$ is the centrifugal force and $A=\pi a^2$ (which note cancels out as $M=\rho A l$).
Requiring $\sigma_{\max}\le\sigma_{\rm break}$ thus implies a maximum spinning frequency of
\begin{equation}
\Omega_{\max} = \frac{1}{l}\sqrt{\frac{8\sigma_{\rm break}}{\rho}} \;,
\label{eq:Omega_max}
\end{equation}
which illustrates the $\Omega_{\max} \sim 1/l$ scaling, with a numerical prefactor depending on the detailed geometry and on the material. 
Inserting $\omega=2\Omega_{\max}$ in Eq.~\eqref{eq:h_opt} gives the strain attainable by a rotor operated at its structural limit,
\begin{equation}
h_{\rm eq}^{\rm cyl, max} = \frac{\pi\,\widetilde{F}(\pi/2)}{1536}\,\frac{G\rho^{2}}{\sigma_{\rm break}}\; l^{2} \left(\frac{l}{r}\right)^{5} \simeq \num{9.5e-3}\,\frac{G\rho^{2}}{\sigma_{\rm break}}\; l^{2} \left(\frac{l}{r}\right)^{5},
\label{eq:h_material}
\end{equation}
which isolates the material figure of merit $\rho^2/\sigma_{\rm break}$ and makes the trade-off explicit. 
At fixed $l/r$ the amplitude of the gravitational signal generated at this limit working-point grows as $h_{\rm eq}^{\rm cyl, max} \sim l^2$ while its frequency falls as $\omega \sim 1/l$, so high frequency and large amplitude cannot be obtained simultaneously.
Density is favored quadratically, whereas a high breaking stress reduces $h_{\rm eq}^{\rm cyl, max}$. Again, it does not weaken the Newtonian field, but it places the signal at a higher frequency where the conversion Eq.~\eqref{eq:heq_general} costs a factor $\omega^{-2}$.
Table~\ref{tab:scalings} collects these dependencies together with those of the torque observables of Sec.~\ref{sec:rotor_detector}, since the equivalent strain and a direct torque measurement respond to the same parameters in markedly different ways.

\renewcommand{\arraystretch}{1.8} 
\setlength{\tabcolsep}{10pt} 

\begin{table}[t]
\centering
\begin{tabular}{llcccccc}
\hline\hline
Varied & Held fixed & $\omega$ & $M$ & $G_{ij}$ & $h_{\rm eq}$ & $N_{0}$ & $\tau_{1}$ \\
\hline
$l$                   & $\omega$, $r$                    & fixed                & $\propto l^{3}$                   & $\propto l^{5}$  & $\propto l^{5}$   & $\propto l^{10}$                  & $\propto l^{-15}$ \\
$l$                   & $\omega$, $\beta$                & fixed                & $\propto l^{3}$                   & unchanged        & unchanged         & $\propto l^{5}$                   & $\propto l^{-5}$ \\
$r$                   & $\omega$, $l$                    & fixed                & unchanged                         & $\propto r^{-5}$ & $\propto r^{-5}$  & $\propto r^{-5}$                  & $\propto r^{10}$ \\
$l$                   & $\beta$, structural limit        & $\propto l^{-1}$     & $\propto l^{3}$                   & unchanged        & $\propto l^{2}$   & $\propto l^{5}$                   & $\propto l^{-5}$ \\
\hline
$\sigma_{\rm break}$  & $\omega$, $\beta$, structural limit                & fixed                & $\propto\sigma_{\rm break}^{3/2}$ & unchanged        & unchanged         & $\propto\sigma_{\rm break}^{5/2}$ & $\propto\sigma_{\rm break}^{-5/2}$ \\
$\rho$                & $\omega$, $l$, $\beta$           & fixed                & $\propto\rho$                     & $\propto\rho$    & $\propto\rho$     & $\propto\rho^{2}$                 & $\propto\rho^{-3}$ \\
$\rho$                & $l$, $\beta$, structural limit   & $\propto\rho^{-1/2}$ & $\propto\rho$                     & $\propto\rho$    & $\propto\rho^{2}$ & $\propto\rho^{2}$                 & $\propto\rho^{-3}$ \\
\hline\hline
\end{tabular}
\caption{\textbf{Scaling of the source and of the two observables.} Each row varies the quantity of the first column, holding those of the second fixed. $\beta\equiv r/l$ is the separation in units of the rotor length and ``structural limit'' means that the rotor is operated at $\omega=2\Omega_{\max}$ of Eq.~\eqref{eq:Omega_max}, which removes $\omega$ from the free parameters. $G_{ij}$ is the tidal amplitude Eq.~\eqref{eq:Gij_general} and $h_{\rm eq}$ the equivalent strain Eq.~\eqref{eq:heq_general}, both at a given direction. $N_{0}$ and $\tau_{1}$ are the gravitational torque Eq.~\eqref{eq:Nmax_cyl} and the time to unit signal-to-noise Eq.~\eqref{eq:tau1_scaling} for two identical rotors in the scheme of Sec.~\ref{sec:rotor_detector}, the latter at fixed bath temperature and damping rate. The material is held fixed except in the lower block, where either $\rho$ or $\sigma_{\rm break}$ are varied. The aspect ratio $\alpha$ is always kept fixed.}
\label{tab:scalings}
\end{table}

\vspace{2mm}
An experimentally validated estimate for $\sigma_{\max}$ is available from the ball-bearing steel spheres of Ref.~\cite{schuck2018}, in which particles of radius $R=\qty{250}{\micro\metre}$ burst at \qty{667}{\kilo\hertz}. 
From this failure frequency, they estimate a breaking stress $\sigma_{\rm break}\simeq\qty{3.4}{\giga\pascal}$, which exceeds the bulk tensile strength of \qty{1.57}{\giga\pascal} assumed in their initial estimate. 
This is expected for millimetre-scale and smaller samples, as the probability of material defects is reduced. 
In general, they find that estimates based on bulk values underpredict the attainable speed of its smallest spheres by up to $46\%$ \cite{schuck2018}, which corresponds to a factor $1.46^{2}=2.1$ in stress and gives \qty{3.3}{\giga\pascal}, consistent with the value quoted above.
Following Ref.~\cite{jose2025}, which assumes the same breaking stress for its NdFeB microrotors since density and geometry are closely similar, we insert $\sigma_{\rm break}=\qty{3.4}{\giga\pascal}$ into Eq.~\eqref{eq:Omega_max} at $\rho=\qty{7430}{\kilo\gram\per\cubic\metre}$, which gives
\begin{equation}
\frac{\Omega_{\max}^{\rm NdFeB}}{2\pi} \simeq \qty{1}{\MHz}\times\frac{\qty{300}{\micro\metre}}{l}.
\label{eq:steel_limit}
\end{equation}

This estimate could be optimistic and in general it is untested. It assumes that a NdFeB microparticle sustains the same stress as hardened bearing steel, whereas the bulk tensile strength of macroscopic sintered samples is of order \qty{80}{\mega\pascal}, so that Eq.~\eqref{eq:steel_limit} presumes a $\sim 40$ times stronger material. This is however not so unrealistic. It is well known that the tensile strength of a sample increases with smaller volume due to smaller size of defects (Weibull law) and with lower temperatures. 
The ultimate theoretical tensile strength of a sample without defects would be actually much higher, given by the Frenkel–Orowan ideal-cleavage estimate to be of the order of $E/10$ where $E$ is the Young modulus \cite{kelly1986}, which would amount to $\sim 15$ GPa for NdFeB \cite{vedrine1991}.
Empirically, the highest stress so far reported by a levitated NdFeB microrotor is \qty{556}{\mega\pascal}, attained at \qty{2.3}{\mega\hertz} with $R=\qty{30}{\micro\metre}$ \cite{jose2025}, an enhancement of about seven over its bulk value. 
The rotor did not fail, however, Ref.~\cite{jose2025} having stopped a factor $\sqrt{3400/556}\simeq2.5$ in frequency below the threshold set by the \qty{3.4}{\giga\pascal} assumed above, to avoid damaging the trap. Both stresses are extracted with the compact-body coefficient, with respect to which the slender-rod coefficient of Eq.~\eqref{eq:Omega_max} is $12\%$ lower in frequency and therefore conservative.
Generating a signal at $f=\qty{6}{\MHz}$, that is $f_{\rm rot}=\qty{3}{\MHz}$, requires $l<\qty{100}{\micro\metre}$ on the optimistic estimate and $l<\qty{41}{\micro\metre}$ on the demonstrated one.

For the equivalent strain, the uncertainty in $\sigma_{\rm break}$ is irrelevant.
At fixed signal frequency and fixed $\beta=r/l$ the structural limit implies $l^{2}=8\sigma_{\rm break}/\rho\Omega^{2}$, the breaking stress cancels from Eq.~\eqref{eq:h_material}, and the attainable strain $h_{\rm eq}^{\rm cyl,max}=\num{9.5e-3}\cdot 8G\rho/\left(\Omega^{2}\beta^{5}\right)$ is set by the density alone. 
For NdFeB it is $h_{\rm eq}^{\rm cyl,max}\simeq\num{3.3e-24}$ at $r=2l$ and $\qty{6}{\MHz}$.
A weaker material does not reduce the strain, it only requires a smaller rotor to reach the same frequency.

This cancellation does not extend to the torque measurement of Sec.~\ref{sec:rotor_detector}, and this should be kept in mind when reading Table~\ref{tab:rotor_detector}, which adopts Eq.~\eqref{eq:steel_limit}. 
There the torque scales as $l^{10}/r^{5}$, that is as $l^{5}$ at fixed $\beta=r/l$, and the torque noise $S_{N}\propto I_{d}$ scales as $l^{5}$ as well, so that $\tau_{1}\propto l^{-5}\propto\sigma_{\rm break}^{-5/2}$ by Eq.~\eqref{eq:tau1_scaling}: the factor $6.1$ between the optimistic and demonstrated breaking stresses is a factor $93$ in integration time. 
On the demonstrated value the \qty{1}{\mega\hertz} row of Table~\ref{tab:rotor_detector} would read $l=\qty{123}{\micro\metre}$, with $\tau_{1}=\qty{4.0}{\year}$ at \qty{4.2}{\kelvin} and \qty{3.5}{\day} at \qty{10}{\milli\kelvin}, rather than \qty{300}{\micro\metre}, \qty{15.9}{\day} and \qty{55}{\minute}.
Establishing the true breaking stress of NdFeB at the hundred-micrometre scale, by spinning a rotor to destruction as in Ref.~\cite{schuck2018}, would help removing the material uncertainty in the torque estimates. 
However, the largest uncertainty lies elsewhere, since $\tau_{1}$ depends far more steeply on the separation ratio than on the material, as $\beta^{10}$ against $\sigma_{\rm break}^{-5/2}$, which makes an exact finite-body calculation at small $\beta$ worthwhile.


\subsection{Near-field gradient from a reciprocating mass}
\label{sec:reciprocating}

In order to generate a time-dependent Newtonian field gradient, an alternative to rotation is linear reciprocation, in which a source mass $M$ is driven back and forth with amplitude $x_{0}$ at angular frequency $\Omega$. 
Because such a mass is clamped to the laboratory and exchanges momentum with its mount, the argument of Sec.~\ref{sec:nearfield_quadrupole} that forbids a time-dependent dipole no longer applies, and the modulated moment is the dipole itself. 
One advantage of this scheme is already seen from the tidal amplitude at distance $r$ along the oscillating direction
\begin{equation}
G_{ij}^{\rm rec} \sim \frac{G M x_0}{r^4},
\label{eq:Grec}
\end{equation}
which shows a decay one power of $r$ more slowly than the quadrupole result Eq.~\eqref{eq:Gijx}. 

For a homogeneous sphere the field can be given in closed form, with no multipole truncation. By Newton's theorem the exterior potential of a sphere of radius $R$ and mass $M=\tfrac{4}{3}\pi\rho R^{3}$ is that of a point mass at its center. 
Taking the oscillation direction to be along $\hat{\mathbf{x}}$, so that the center sits at $\mathbf{s}(t)=x_{0}\cos(\Omega t)\,\hat{\mathbf{x}}$, the definitions in Eq.~\eqref{eq:g_Gij_defs} give
\begin{equation}
\Phi(\mathbf{r},t) = -\frac{GM}{|\mathbf{r}-\mathbf{s}(t)|}, \qquad
G_{ij}(\mathbf{r},t) = \frac{GM}{d(t)^{3}}\left(3N_{i}N_{j}-\delta_{ij}\right),
\label{eq:Grec_exact}
\end{equation}
with $d(t)\equiv|\mathbf{r}-\mathbf{s}(t)|$ and $\mathbf{N}\equiv(\mathbf{r}-\mathbf{s})/d(t)$. 
Equation~\eqref{eq:Grec_exact} is exact for $d(t)>R$. 
This is in contrast with the rotor, for which the multipole expansion of Sec.~\ref{sec:nearfield_quadrupole} is unavoidable because no exterior equivalence of this kind exists.

We evaluate the field at the same two points $P_x$ and $P_z$ used for the rotor, at a distance $r>R+x_{0}$ from the origin. 
Along the drive axis, $\mathbf{r}=(r,0,0)\equiv P_{x}$, one has $\mathbf{N}=\hat{\mathbf{x}}$ and $d(t)=r-x_{0}\cos(\Omega t)$, so that
\begin{equation}
G_{ij}(P_{x},t) = \frac{GM}{d(t)^{3}}
\begin{pmatrix} 2 & 0 & 0 \\ 0 & -1 & 0 \\ 0 & 0 & -1 \end{pmatrix}_{\! ij} .
\label{eq:Grec_axis}
\end{equation}
The tensor structure is thus fixed and only the scalar prefactor oscillates. 
The requirement $r>R+x_{0}$ introduced above guarantees $d(t)\ge r-x_{0}>R>0$ throughout the oscillation cycle, so that $P_x$ is always outside the moving body at any time $t$, and it implies $\varepsilon\equiv x_{0}/r<1$, which is the condition for the following expansion of $1/d(t)$ to converge. 
Resolving the resulting powers of $\cos(\Omega t)$ into harmonics,
\begin{equation}
\frac{GM}{d^{3}(t)} = \frac{GM}{r^{3}}\Big[\underbrace{1+3\varepsilon^{2}}_{\rm static}
+\underbrace{\left(3\varepsilon+\tfrac{15}{2}\varepsilon^{3}\right)\cos(\Omega t)}_{\rm signal}
+\underbrace{3\varepsilon^{2}\cos (2\Omega t) + \tfrac{5}{2}\varepsilon^{3}\cos (3\Omega t)}_{\rm harmonics}+\mathcal{O}(\varepsilon^{4})\Big],
\label{eq:Grec_harmonics}
\end{equation}
whose fundamental has amplitude $G_{xx}^{(\Omega)}=6GMx_{0}/r^{4}$. The static term is the monopole gradient discussed below Eq.~\eqref{eq:Gij_full} and is again neglected. 
Two differences from the rotor are already visible. 
First, it is interesting to note that the rotor field Eq.~\eqref{eq:Gijx} is exactly monochromatic at $2\Omega$, because $Q_{ij}(t)$ is exactly sinusoidal, whereas a reciprocating source gives an intrinsically anharmonic signal, with its $n$-th harmonic entering at order $\varepsilon^{n}$. 
More importantly, its strongest harmonic appears at the drive frequency itself rather than at twice it, so that the separation which protects the rotor against drive-synchronous systematics is lost. Any residual coupling of the actuator to the test mass falls in the signal band and is spectrally indistinguishable from it. 
For $x_{0}=R=l/2$ and $r=2l$, i.e. the working point considered later in Eq.~\eqref{eq:rec_vs_rot}, one has $\varepsilon=1/4$ and the second harmonic reaches $\sim 22\%$ of the fundamental, so that a reciprocating source driven at $\Omega$ deposits a gradient at $2\Omega$ of about a fifth of its own signal amplitude, precisely in the band where a rotor of the same frequency would be measured.

\begin{figure}
    \centering
    \includegraphics[width=0.4\linewidth, valign=c]{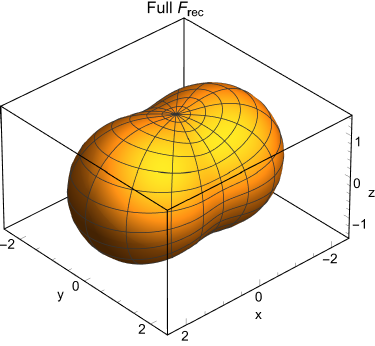}
    \includegraphics[width=0.45\linewidth, valign=c]{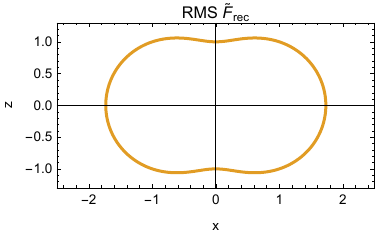}
    \caption{\textbf{Angular pattern of the near-field tidal amplitude for a reciprocating mass.} Normalized amplitude of the field-gradient tensor at fixed distance from the reciprocating sphere. Left: the instantaneous pattern $F_{\rm rec}(\theta,\varphi,t)$ of Eq.~\eqref{eq:Grec_norm}, shown at time $t=0$. The two lobes reflect the $m=1$ structure of the reciprocating mass. The pattern reaches $F_{\rm rec}=\sqrt{6}$ along the lobe axes and $F_{\rm rec}=\sqrt{2}$ between them, so the azimuthal contrast is a factor $1.73$ and there are no nodes. Right: the $\varphi=0$ section of the cycle-averaged pattern $\widetilde{F}_{\rm rec}$ of Eq.~\eqref{eq:Frec}, obtained by averaging $F^{2}_{\rm rec}$ over one oscillation period. Unlike the rotor, the cycle average does not remove the azimuthal dependence here, since $t$ and $\varphi$ do not combine into a single phase in Eq.~\eqref{eq:Grec_norm}. The pattern is instead axially symmetric about the drive axis $\hat{\mathbf{x}}$, so the full three-dimensional surface is the figure of revolution of the curve shown about the horizontal axis. The amplitude grows monotonically from $\widetilde{F}_{\rm rec}=1$ on the vertical $z-$axis to $\widetilde{F}_{\rm rec}=\sqrt{3}$ on the horizontal $x-$axis, which is thus the preferred location for a detector.}
    \label{figPatternRecip}
\end{figure}

For a field point $\mathbf{r}$ at distance $r=|\mathbf{r}| > R+x_{0}$ and arbitrary direction $\mathbf{n}=\mathbf{r}/r$, it is sufficient to work to first order in $x_{0}/r$. 
Expanding the kernel in the displacement exactly as in Eq.~\eqref{eq:greens_expand}, but with $\mathbf{s}(t)$ in place of the source coordinate and with no integration to perform (since the sphere is effectively a point mass outside itself), we get
\begin{equation}
\frac{1}{|\mathbf{r}-\mathbf{s}(t)|} = \frac{1}{r} + \frac{s_{k}(t)\,r_{k}}{r^{3}} + \mathcal{O}\left( ( x_0/r )^2 \right) = \frac{1}{r} + x_{0}\cos(\Omega t)\,\frac{\hat{x}_{k}r_{k}}{r^{3}} + \mathcal{O}\left( ( x_0/r )^2 \right) ,
\label{eq:rec_expand}
\end{equation}
the neglected terms being those responsible for the static shift and the second harmonic in Eq.~\eqref{eq:Grec_harmonics}. The entire time dependence has factored out of the geometry into the scalar $x_{0}\cos(\Omega t)$, which is why the first-order field is exactly monochromatic although the full field in Eq.~\eqref{eq:Grec_axis} is not. 
Differentiating twice with respect to the field point, with
$\partial_{i}\partial_{j}\partial_{k}\left(1/r\right)=-3\left[5n_{i}n_{j}n_{k}-\left(n_{i}\delta_{jk}+n_{j}\delta_{ik}+n_{k}\delta_{ij}\right)\right]/r^{4}$, the oscillating part of the tidal tensor is
\begin{equation}
G^{\rm rec}_{ij}(\mathbf{r},t) = \frac{3GMx_{0}}{r^{4}}\cos(\Omega t) \,
\Big[\,5\,u\,n_{i}n_{j}-\left(\hat{x}_{i}n_{j}+\hat{x}_{j}n_{i}+u\,\delta_{ij}\right)\Big],
\qquad u\equiv \mathbf{n}\cdot\hat{\mathbf{x}}=\sin\theta\cos\varphi ,
\label{eq:Grec_general}
\end{equation}
where $u$ is the cosine of the angle between the field direction and the drive axis.
Equation~\eqref{eq:Grec_general} is manifestly symmetric and trace free, since $n_{i}n_{i}=1$, $\hat{x}_{i}n_{i}=u$ and $\delta_{ii}=3$ give $5u-(u+u+3u)=0$. 
At $P_{x}$, where $u=1$, it reduces to $\frac{3GM x_0}{r^4}\cos(\Omega t) \mathrm{diag}(2,-1,-1)$, recovering the fundamental term of Eq.~\eqref{eq:Grec_harmonics}. At $P_{z}=(0,0,r)$, transverse to the drive, $u$ vanishes and only the shear component survives,
\begin{equation}
G^{\rm rec}_{xz}(P_{z},t)=G^{\rm rec}_{zx}(P_{z},t)=-\frac{3GMx_{0}}{r^{4}}\cos(\Omega t) ,
\label{eq:Grec_Pz}
\end{equation}
smaller by a factor two than the largest component along the oscillation axis.

Contracting Eq.~\eqref{eq:Grec_general} gives the squared Frobenius norm for the tidal tensor
\begin{equation}
G^{\rm rec}_{ij}G^{\rm rec}_{ij} = 9\,\frac{G^{2}M^{2}x_{0}^{2}}{r^{8}}\,F_{\rm rec}^{2}(\theta,\varphi,t),
\qquad F_{\rm rec}(\theta,\varphi,t) \equiv \left|\cos(\Omega t)\right| \sqrt{2} \sqrt{1+2\sin^{2}\theta\cos^{2}\varphi} ,
\label{eq:Grec_norm}
\end{equation}
to be compared with Eq.~\eqref{eq:Ginvariant}. 
The factor $\sqrt{2}$ normalizes $F_{\rm rec}(\theta,\varphi,t)$ to unity at $P_{z}$ when averaged over a cycle.
In fact, we have
\begin{equation}
\widetilde{F}_{\rm rec}(\theta,\varphi) \equiv \left\langle F_{\rm rec}^{2}(\theta,\varphi,t) \right\rangle_{t}^{1/2}
= \sqrt{1+2\sin^{2}\theta\cos^{2}\varphi} ,
\label{eq:Frec}
\end{equation}
with $\widetilde{F}_{\rm rec}(P_{z})=1$ and $\widetilde{F}_{\rm rec}(P_{x})=\sqrt{3}\simeq1.73$, so that the pattern is less anisotropic than that of the rotor, which according to Eq.~\eqref{eq:pattern} varies by a factor $4.65$ over the same two points, see Fig.~\ref{figPatternRecip}. 
As done before, equating Eqs.~\eqref{eq:Grec_norm} to Eq.~\eqref{eq:GGgw} in the root-mean-square sense, the equivalent strain of a reciprocating sphere is
\begin{equation}
h_{\rm eq}^{\rm rec}(\theta,\varphi) = 3\sqrt{2}\,\frac{G M x_{0}}{\Omega^{2}r^{4}}\,\widetilde{F}_{\rm rec}(\theta,\varphi) ,
\label{eq:heq_rec}
\end{equation}
falling one power of $r$ more slowly than the quadrupole result Eq.~\eqref{eq:heq_general}, as expected for a dipole source.

The two schemes can now be compared. A sphere of diameter $l=2R$ has exactly the mass $M=\pi\rho l^{3}/6$ of the optimal cylinder of the same length, so that at equal density, equal overall size, equal separation $r=2l$ and equal signal frequency $\omega$ (the rotor spinning at $\omega/2$ and the sphere being driven at $\omega$), each evaluated on its own most favourable axis, Eqs.~\eqref{eq:heq_rec} and Eq.~\eqref{eq:h_opt} give
\begin{equation}
\frac{h_{\rm eq}^{\rm rec}(P_{x})}{h_{\rm eq}^{\rm cyl}(\pi/2)} \simeq 25.3\,\frac{x_{0}}{l} \;,
\label{eq:rec_vs_rot}
\end{equation}
favoring reciprocation by an order of magnitude once $x_{0}\sim l/2$.
Whether this advantage is accessible, however, depends on the frequency of interest.
A body reciprocating with peak acceleration $\omega^{2}x_{0}$ sustains an internal stress of order $\rho\,\omega^{2}x_{0}\,l/2$, so that the failure criterion used for the rotor limits the amplitude to $x_{0,\rm max} \lesssim 2\sigma_{\rm break}/(\rho\omega^{2}l)$. Inserting this into Eq.~\eqref{eq:rec_vs_rot} and using $\rho\,\omega^{2}l^{2}=32\,\sigma_{\rm break}$ for a rotor at its own structural limit, which is Eq.~\eqref{eq:Omega_max} written in terms of the signal frequency $\omega=2\Omega_{\max}$, leaves
\begin{equation}
\frac{h_{\rm eq}^{\rm rec}}{h_{\rm eq}^{\rm cyl}} \lesssim \frac{25.3}{16} \simeq 1.6 ,
\label{eq:rec_vs_rot_limited}
\end{equation}
independent of size, frequency and material. Concretely, at $f=\qty{6}{\MHz}$ and $l=\qty{100}{\micro\metre}$ the admissible amplitude is only $x_{0}\simeq\qty{6}{\micro\metre}$; driving the same sphere at $x_{0}=l/2$ would demand a peak acceleration of $\num{7e10}\,\mathrm{m/s^{2}}$ and internal stresses of order \qty{26}{\giga\pascal}, roughly an order of magnitude beyond any breaking limit.
 
Reciprocation is therefore particularly attractive at low frequencies, where large displacements can be achieved with comparatively modest accelerations, the drive waveform can be tailored for systematic checks, and the translational motion can, in principle, be prepared in nonclassical states for experiments probing gravitationally mediated quantum correlations. At MHz frequencies and above, rotation offers a more favourable route to strong gravitational modulation: the source can sustain rapid motion without periodically accelerating its centre of mass, while the gravitational signal at twice the rotation frequency is spectrally separated from the leading mechanical disturbances associated with rotor imbalance.

\section{Detection with a second spinning rotor}
\label{sec:rotor_detector}

The field gradient derived in Sec.~\ref{sec:cylinder} oscillates at twice the rotation rate of the source and is confined to the near-field zone within a few source diameters. 
Detecting it requires a device that couples to a tidal tensor rather than to a force, that can be placed at a separation comparable to the size of the source, and that remains sensitive in the desired frequency range. 
A natural detector to consider is a second rotor, whose quadrupole couples to the gradient through a torque and whose own rotation downconverts the signal to a band of a few hertz. 
The scheme is entirely classical, which has the advantage that its sensitivity and systematic control can be assessed with existing numbers. Preparing a rotor in nonclassical states is a natural extension that we leave to a future work.

\subsection{Gravitational torque}

The gravitational torque on a body about its own center of mass is given by
\begin{equation}
N_i = \varepsilon_{ijk}\int \,d^{3}x\, \rho_{\rm det}(\mathbf{x}) \, x_j\, g_k(\mathbf{x}), \qquad g_k(\mathbf{x}) \simeq g_k(0) + G_{kl}\,x_l ,
\label{eq:torque_def}
\end{equation}
with the field linearized about the detector center, which is legitimate to the extent that the detector is small compared with $r$. The uniform part $g_k(0)$ exerts no torque about the center of mass, since $\int d^{3}x\,\rho_{\rm det}(\mathbf{x})\,x_j=0$ there. 
Carrying out the remaining integral gives $\varepsilon_{ijk}G_{kl}\int d^{3}x \,\rho_{\rm det}(\mathbf{x})\, x_jx_l$, and inserting $\int d^{3}x \,\rho_{\rm det}(\mathbf{x})\, x_jx_l = \tfrac{1}{3}\left(Q^{\rm det}_{jl} + \delta_{jl}\int|\mathbf{x}|^{2}\rho_{\rm det}(\mathbf{x})\,d^{3}x\right)$ from the definition Eq.~\eqref{eq:Qij_def} of the detector quadrupole $Q^{\rm det}_{ij}$, the trace term drops against the antisymmetry of $\varepsilon_{ijk}$ and the symmetry of $G_{kl}$, leaving
\begin{equation}
N_i = \frac{1}{3}\,\varepsilon_{ijk}\, G_{kl}\, Q^{\rm det}_{jl}(t).
\label{eq:torqueG}
\end{equation}
Only the quadrupole of the detector enters at this order, exactly as only the quadrupole of the source generates the field, so the coupling is quadrupole to quadrupole and the configuration is symmetric between the two bodies. Higher detector moments couple to higher gradients of the source field and are suppressed by further powers of $l_{\rm det}/r$, with $l_{\rm det}$ the detector length, as in Sec.~\ref{sec:heq}.

\subsubsection{Coaxial and side-by-side configurations}

Two rotors spinning about parallel axes, both taken along $z$, admit distinct placements. 
We discuss two relevant scenarios, one coaxial and one side-by-side.
In the coaxial configuration the detector lies on the spin axis of the source at $\mathbf{r}=(0,0,r)$ and samples the axial gradient Eq.~\eqref{eq:Gijz}. In the side-by-side configuration it is displaced perpendicular to the spin axis at $\mathbf{r}=(r,0,0)$ and samples the equatorial gradient Eq.~\eqref{eq:Gijx}. In both cases the detector quadrupole is
\begin{equation}
Q^{\rm det}_{ij}(t) = Q^{\rm det} \begin{pmatrix} \cos 2\left(\Omega_{d}t+\phi_{0}\right) & \sin 2\left(\Omega_{d}t+\phi_{0}\right) & 0 \\ \sin 2\left(\Omega_{d}t+\phi_{0}\right) & -\cos 2\left(\Omega_{d}t+\phi_{0}\right) & 0 \\ 0 & 0 & 0 \end{pmatrix}_{ij},
\label{eq:Qdetector}
\end{equation}
with $\Omega_{d}$ the detector angular velocity, negative for counter-rotation, and $\phi_{0}$ the relative phase. Since $Q^{\rm det}_{ij}$ vanishes whenever either index equals $z$, and both gradients Eq.~\eqref{eq:Gijz} and Eq.~\eqref{eq:Gijx} satisfy $G_{zx}=G_{zy}=0$, the transverse components of Eq.~\eqref{eq:torqueG} vanish identically in both geometries and the torque is purely axial.

Inserting Eq.~\eqref{eq:Qdetector} together with Eq.~\eqref{eq:Gijz} or Eq.~\eqref{eq:Gijx} into Eq.~\eqref{eq:torqueG}, and introducing the difference and sum frequencies
\begin{equation}
\delta \equiv \Omega - \Omega_{d}, \qquad \Sigma \equiv \Omega + \Omega_{d},
\label{eq:delta_sigma}
\end{equation}
both cases take the common form
\begin{equation}
N_{z}(t) = \frac{G\,Q\,Q^{\rm det}}{r^{5}}\left[\kappa_{-}\sin\left(2\delta t - 2\phi_{0}\right) + \kappa_{+}\sin\left(2\Sigma t + 2\phi_{0}\right)\right],
\label{eq:torque}
\end{equation}
with the geometrical coefficients
\begin{equation}
\left(\kappa_{-},\,\kappa_{+}\right) = \begin{cases} \left(\tfrac{2}{3},\ 0\right) & \text{coaxial}, \\[4pt] \left(\tfrac{1}{4},\ -\tfrac{35}{12}\right) & \text{side-by-side}. \end{cases}
\label{eq:kappas}
\end{equation}
The origin of the difference is the polarization of the source gradient. 
The axial gradient carries a single circular polarization, so only the difference-frequency term is driven and the sum-frequency term is exactly annulled.
The equatorial gradient is elliptically polarized, contains both circular components, and therefore drives both terms, the sum-frequency coupling being the stronger by an order of magnitude.

A slowly varying, or even static, torque is obtained by bringing one argument of Eq.~\eqref{eq:torque} to rest. 
Either geometry can be operated with the two rotors co-rotating, $\Omega_{d}=+\Omega$, which brings $\delta$ to rest and selects $\kappa_{-}$, or counter-rotating, $\Omega_{d}=-\Omega$, which brings $\Sigma$ to rest and selects $\kappa_{+}$. The four combinations are therefore the four entries of Eq.~\eqref{eq:kappas}. The coaxial geometry is useful only co-rotating, where it delivers the amplitude $\tfrac{2}{3}GQQ^{\rm det}/r^{5}$, since counter-rotation selects $\kappa_{+}=0$ and thus yields no static torque at all. 
The side-by-side geometry is best counter-rotating, delivering $\tfrac{35}{12}GQQ^{\rm det}/r^{5}$, larger by a factor $4.4$ at equal separation and hence shorter in integration time by a factor $19$ (see later Eq.~\eqref{eq:SNR}). 
Co-rotating would retain here only $\kappa_{-}=1/4$. 
At a given separation the counter-rotating side-by-side pair is therefore the strongest of the four configurations.
Whenever a static torque is present it reaches its extrema at $\phi_{0}=\pi/4+n\pi/2$ and vanishes at $\phi_{0}=n\pi/2$, the four extrema per turn reflecting the quadrupolar character of both bodies. 
In the side-by-side case the term not brought to rest remains, at $4\Omega$, and might be filtered by the low-pass response of the readout, while in the coaxial case it is absent altogether, since $\kappa_{+}=0$.

The comparison above is not settled by Eq.~\eqref{eq:kappas} alone, because the fifth-power scaling makes the minimum achievable separation as important as the coupling coefficient. 
A cylinder of length $l$ and radius $a$ spinning about a transverse axis sweeps an envelope of radius $\sqrt{l^{2}/4+a^{2}}$ and height $2a$. Two identical rotors therefore clear at $r_{\min}=2a=2\alpha l$ when stacked along the common spin axis, the envelopes being then separated vertically, and at $r_{\min}=l\sqrt{1+4\alpha^{2}}$ when placed side by side, the envelopes being separated horizontally. 
For the optimal aspect ratio $\alpha=1/\sqrt{6}$ these are $0.816\,l$ and $1.291\,l$ respectively, and evaluating $\kappa\left(l/r_{\min}\right)^{5}$, with $\kappa$ the optimal coefficient in the considered configuration, gives $1.84$ for the stacked co-rotating pair ($\kappa_{-}=2/3$) against $0.81$ for the side-by-side counter-rotating pair ($\kappa_{+}=35/12$).
The larger coupling coefficient of the side-by-side geometry is thus suppressed by its larger clearance, and at contact the stacked configuration is superior by a factor $2.3$.

Interestingly, once the minimum separation is determined by these geometrical arguments, the aspect ratio of the rotor can be optimized further in order to maximize the signal. 
The optimum is then shifted away from $\alpha=1/\sqrt{6}$, and it differs between the two configurations. 
Using Eq.~\eqref{eq:Qcyl}, the figure of merit at contact is
\begin{equation}
\kappa\,\frac{Q}{r_{\min}^{5}} = \frac{\pi\rho}{8}\times \begin{cases} \dfrac{2}{3}\,\dfrac{1-3\alpha^{2}}{32\,\alpha^{3}} & \text{coaxial}, \\[10pt] \dfrac{35}{12}\,\dfrac{\alpha^{2}\left(1-3\alpha^{2}\right)}{\left(1+4\alpha^{2}\right)^{5/2}} & \text{side by side}. \end{cases}
\label{eq:contact_fom}
\end{equation}
The side-by-side expression is maximized at $\alpha=\sqrt{1-\sqrt{30}/6}\simeq0.295$, the optimum being shifted below the value $1/\sqrt{6}$ of Sec.~\ref{sec:cylinder} because the clearance itself grows with the rotor thickness.
The coaxial expression instead increases monotonically as the rotor is made thinner, since the quadrupole falls only as $\alpha^{2}$ while the closer approach gains $\alpha^{-5}$.
Apart from their common zero at $\alpha=1/\sqrt{3}$, the two expressions cross only at $\alpha=0.557$.
Since rotational stability requires the spin axis to carry the largest moment of inertia and hence $\alpha<1/\sqrt{3}\simeq0.577$, the stacked geometry is superior over essentially the entire range of stable aspect ratios. The optimum is then set not by Eq.~\eqref{eq:Qcyl} but by the levitation stiffness and by the elastic deformation of a slender rod at full speed. These contact estimates must in any case be read as indicative rather than quantitative, since $r_{\min}=2\alpha l$ is smaller than $l$ for any $\alpha<1/2$: already at $\alpha=1/\sqrt{6}$ the expansion parameter is $(l/r_{\min})^{2}=1.5$, so that the quadrupole expansion of Sec.~\ref{sec:nearfield_quadrupole} has broken down and the exact Newtonian interaction of the two finite bodies would have to be computed. The divergence of Eq.~\eqref{eq:contact_fom} as $\alpha\to 0$ is in particular just artificial.

Two further considerations favor the coaxial geometry.
First, it preserves the exact correspondence with a circularly polarized gravitational wave established in Sec.~\ref{sec:heq}, so it admits an unambiguous interpretation as a near-field analogue of wave detection, whereas the side-by-side geometry is a comparison of magnitudes alone.
Second, the stray magnetic field, which at equal separation is weakest along the spin axis. A dipole rotating in the $xy$ plane presents its equatorial field $\mu_{0}m/4\pi r^{3}$ at every point of the $z$ axis, whereas a point in the plane of rotation sees a field oscillating between this value and twice it as the dipole sweeps past.
If each geometry is used at its own geometrical clearance, the advantage appears reversed. In units of $\mu_{0}m/4\pi l^{3}$, the stray field at $r_{\min}$ is $1.84$ for the coaxial configuration and $0.93$ for the side by side one, the closer coaxial approach outweighing the polar-to-equatorial factor two through the $1/r^{3}$ dependence.
This does not, however, necessarily disadvantage the coaxial choice, since the magnetic coupling can be rejected equally well in both configurations by reversing one of the rotors by half-turn, as discussed in Sec.~\ref{sec:syscoup}. 

In what follows we thus focus on the coaxial configuration, for which $\kappa_{-}=2/3$ and $\kappa_{+}=0$, and we indicate where the side-by-side alternative would alter the conclusions. 
The side-by-side alternative is preferable only if a constraint external to the rotors, such as the thickness of a shielding divide, imposes a clearance along the spin axis exceeding $0.96\,l$, some $18\%$ beyond the contact value $2a=0.816\,l$.

\subsection{The detector rotor as a mixer}

Equation \eqref{eq:torque} is a central result of this section. 
As we can see, while the gravitational field oscillates at $2\Omega$, e.g. in the megahertz range, in the coaxial configuration the observable torque oscillates at $2\delta$. 
The torque frequency is thus set by the experimenter through the choice of $\Omega_{d}$ and may be made arbitrarily small. In this sense, the detector rotor performs a mechanical signal downconversion, acting as a phase-coherent mixer whose local oscillator is its own rotation.
A high-frequency gravitational signal can be down-converted to a band of a few hertz, where torque detection is most sensitive. 
The sum-frequency line at $2\Sigma$, present only for the side-by-side geometry, lies near $4\Omega$ and can be removed by low-pass filtering on the readout.

This freedom in the choice of $\delta$ is not unlimited, because the quantity actually recorded is an angle rather than a torque. 
A freely spinning rotor driven at $2\delta$ librates with amplitude $N_{0}/I_{d}\left(2\delta\right)^{2}$, so the angular signal falls as the inverse square of the detuning. 
The thermal torque noise, Eq.~\eqref{eq:SNth} below, is also converted into an angle by the same factor, so the ratio of the two is independent of $\delta$, but the angle noise of the readout is not converted at all and it is what determines the usable band. 
Equating the two gives a maximum detuning $2\delta\lesssim\left(S_{N}/I_{d}^{2}S_{\theta}\right)^{1/4}$, with $S_{\theta}$ the readout angle noise spectral density, which for the \qty{1}{\mega\hertz} row of Table~\ref{tab:rotor_detector} and a reference $\sqrt{S_{\theta}}=\qty{e-9}{\radian\rtHz}$ amounts to a fraction of a hertz, widening only as $S_{\theta}^{-1/4}$. 
The detector is in this sense band limited about zero frequency, similarly to a resonant antenna that is band limited about its mechanical resonance, the essential difference being that the center of the band is set by $\Omega_{d}$ and is continuously tunable rather than fixed by construction. 
The same expression appears again in the discussion in Sec.~\ref{sec:gw_antenna} as Eq.~\eqref{eq:gw_band}, where it sets the instantaneous bandwidth of a rotor used as a gravitational-wave antenna.

The synchronous point $\delta=0$ is not the optimal operating point, despite giving the largest instantaneous torque. A static torque must be read out at zero frequency, where the apparatus is mostly subject to background noise. 
The best choice is a small nonzero detuning, placing the signal at $2\delta$ in a band chosen to be free of mechanical resonances and above the $1/f$ noise limit of the readout, which for the parameters above means a fraction of a hertz to a few hertz. 
This is a narrow window, and the magnetic leakage requirement Eq.~\eqref{eq:shield_requirement}, which relaxes as $\delta^{-1}$, pushes against its upper end, so that improving $S_{\theta}$ improves both bandwidth and shielding tolerance. 
The signal is then a harmonic of known frequency and of phase locked to the independently measured angular positions of the two rotors, which permits effective filtering. 
Because the thermal torque noise Eq.~\eqref{eq:SNth} is white, nothing is lost in sensitivity by this band displacement, and a significant advantage is gained in filtering. Reversing the sign of $\delta$, or stepping it between values, provides a further discriminant, since the gravitational signal follows while most instrumental artifacts do not.

\subsubsection{Thermal noise limit and integration time}

Here, we consider a freely spinning rotor and we assume that the detection is thermal-noise limited at the signal frequency.
For a detector rotor with moment of inertia $I_{d}$ about the spin axis and damping rate $\gamma$, defined by $\dot{\Omega}_{d}=-\gamma\Omega_{d}$ so that the free spin-down time is $1/\gamma$, the fluctuation-dissipation theorem gives the single-sided torque noise spectral density
\begin{equation}
S_{N} = 4k_{B}T I_{d}\gamma,
\label{eq:SNth}
\end{equation}
with $T$ the temperature of the bath responsible for the dissipation, in practice residual gas and, for a magnetically levitated rotor, hysteretic and eddy-current losses in the superconductor and in nearby conductors. 
Writing the torque amplitude of Eq.~\eqref{eq:torque} as
\begin{equation}
N_{0} = \left|\kappa\right| \frac{G\,Q\,Q^{\rm det}}{r^{5}},
\label{eq:N0}
\end{equation}
with $\kappa$ depending on the chosen configuration, matched filtering of a sinusoid of known frequency and phase over an integration time $\tau$ gives
\begin{equation}
\mathrm{SNR} = N_{0}\sqrt{\frac{\tau}{2S_{N}}}, \qquad\Rightarrow\qquad \tau_{1} = \frac{2S_{N}}{N_{0}^{2}} \quad \text{for SNR$=1$},
\label{eq:SNR}
\end{equation}
the factor of two arising from the mean square of a sinusoid. This would be absent for a measurement held at exactly $\delta=0$, but we retain it throughout, since a strictly static readout would be subject to the low-frequency drifts that motivated modulation in the first place.

Taking two identical cylinders of length $l$ and aspect ratio $\alpha=1/\sqrt{6}$, for which Eq.~\eqref{eq:Qmax} gives $Q=Q^{\rm det}=\pi\rho l^{5}/96$ and the moment of inertia about the spin axis is $I_{d}=Ml^{2}/8=\pi\rho l^{5}/48=2Q$, the torque takes the compact form
\begin{equation}
N_{0} = \left|\kappa\right| \frac{G I_{d}^{2}}{4r^{5}} = \left|\kappa\right| \frac{\pi^{2}G\rho^{2}l^{10}}{9216\,r^{5}},
\label{eq:Nmax_cyl}
\end{equation}
so that at fixed ratio $\beta = r/l$ the torque grows as the fifth power of the rotor size, $N_{0}=\left|\kappa\right|\pi^{2}G\rho^{2}l^{5}/9216\,\beta^{5}$. Since $S_{N}\propto I_{d}\propto l^{5}$, the time to achieve unit signal-to-noise is then
\begin{equation}
\tau_{1} = \frac{2S_{N}}{N_{0}^{2}} = \frac{128\,k_{B}T\gamma}{\kappa^{2}G^{2}}\,\frac{r^{10}}{I_{d}^{3}} = \frac{128\cdot 48^{3}}{\pi^{3}}\,\frac{k_{B}T\gamma}{\kappa^{2}G^{2}\rho^{3}}\,\frac{\beta^{10}}{l^{5}},
\label{eq:tau1_scaling}
\end{equation}
falling as the fifth power of the rotor size and rising as the tenth power of the separation ratio. The dependence on $\beta$ is the dominant one by a wide margin. Operating each rotor at its structural limit, where $l\propto 1/\Omega_{\max}$ by Eq.~\eqref{eq:Omega_max}, converts Eq.~\eqref{eq:tau1_scaling} at fixed $\beta$ into $\tau_{1}\propto \Omega_{\max}^{5}$, so that each order of magnitude in frequency costs five orders of magnitude in measurement time.

Table~\ref{tab:rotor_detector} evaluates these expressions for the coaxial configuration with NdFeB rotors at $\beta=2$ and $\gamma=\qty{e-7}{\per\second}$, corresponding to a free spin-down time of about four months. A damping rate $\gamma=\qty{4.75e-7}{\per\second}$ has already been reported on a Meissner-levitated microrotor at \qty{4.2}{\kelvin} \cite{jose2025} and more recent unpublished measurements are approaching $\gamma=\qty{e-7}{\per\second}$. Thus the value assumed here appears realistic or even conservative, given that further improvements may be expected at lower temperatures.
The temperature is the more demanding assumption, and we therefore quote $\tau_{1}$ at two values of it. The first, \qty{4.2}{\kelvin}, is the temperature at which the mentioned damping rate was demonstrated. 
The second, \qty{10}{\milli\kelvin}, is a dilution-refrigerator value, considerably harder to reach for a levitated rotor than for a clamped one. The body is thermally isolated by construction and can exchange heat only through the residual gas and through radiation, so its internal temperature is set by the balance against the heat deposited by the trap and the readout rather than by the temperature of the cryostat. 
Since $S_{N}\propto T\gamma$, the two columns differ by the fixed factor $420$ and may be rescaled to any other temperature at will. Each row takes the rotor length at the structural limit for the quoted rotation frequency following Eq.~\eqref{eq:steel_limit}.

\renewcommand{\arraystretch}{1.8} 
\setlength{\tabcolsep}{12pt} 

\begin{table}[t]
\centering
\begin{tabular}{ccccccc}
\hline\hline
$f_{\rm rot}=\Omega_{\max}/2\pi$ & $l$ & $M$ & $I_{d}$ & $N_{0}$ & $\tau_{1}^{\qty{4.2}{\kelvin}}$ & $\tau_{1}^{\qty{10}{\milli\kelvin}}$ \\
\hline
\qty{100}{\kilo\hertz} & \qty{3.00}{\milli\metre} & \qty{105}{\milli\gram} & \qty{1.2e-10}{\kilo\gram\metre\squared} & \qty{2.0e-20}{\newton\metre} & \qty{13.7}{\second} & \qty{33}{\milli\second} \\
\qty{1}{\mega\hertz} & \qty{300}{\micro\metre} & \qty{105}{\micro\gram} & \qty{1.2e-15}{\kilo\gram\metre\squared} & \qty{2.0e-25}{\newton\metre} & \qty{15.9}{\day} & \qty{55}{\minute} \\
\qty{10}{\mega\hertz} & \qty{30.0}{\micro\metre} & \qty{105}{\nano\gram} & \qty{1.2e-20}{\kilo\gram\metre\squared} & \qty{2.0e-30}{\newton\metre} & \qty{4350}{\year} & \qty{10.4}{\year} \\
\hline\hline
\end{tabular}
\caption{Gravitational torque between two identical optimal NdFeB cylinders ($\alpha=1/\sqrt{6}$) in the coaxial configuration at separation $r=2l$, each spinning at the structural limit for the quoted rotation frequency, and the integration time required for unit signal-to-noise at $\gamma=\qty{e-7}{\per\second}$. Since $\tau_{1}\propto T$ by Eq.~\eqref{eq:tau1_scaling}, the integration time columns differ by $420$, the temperature ratio. 
}
\label{tab:rotor_detector}
\end{table}

The entries of Table~\ref{tab:rotor_detector} are conservative in one important respect. The value $\beta=2$ is imposed by the validity of the quadrupole expansion rather than by the geometry, the two rotors coming into contact at $\beta=0.816$ in the coaxial configuration and at $\beta=1.291$ in the side-by-side one, as discussed in Sec.~\ref{sec:rotor_detector}. Through the tenth power in Eq.~\eqref{eq:tau1_scaling}, closing the gap from $\beta=2$ to $\beta=1$ shortens $\tau_{1}$ by three orders of magnitude, and reaching contact in the coaxial geometry would shorten it by a further factor of nearly eight. The \qty{10}{\mega\hertz} entry would then fall from \qty{10.4}{\year} to about half a day at \qty{10}{\milli\kelvin}, and from \qty{4350}{\year} to some seven months at \qty{4.2}{\kelvin}. 
Exploiting this requires abandoning the multipole expansion and numerically computing the exact Newtonian interaction of the two finite bodies.

Within the quadrupole treatment, the fifth-power scaling of Eq.~\eqref{eq:tau1_scaling} confines the useful range of this detection scheme to between roughly \qty{100}{\kilo\hertz} and a few megahertz, the lower end being set not by the noise, which continues to fall, but by the practical difficulty of levitating a rotor of millimeter size and by low-frequency drifts. 
At \qty{1}{\mega\hertz} rotation, corresponding to a gravitational signal at \qty{2}{\mega\hertz}, some \qty{55}{\minute} of integration suffice in principle at \qty{10}{\milli\kelvin}, and \qty{15.9}{\day} at \qty{4.2}{\kelvin}.
This is a striking conclusion given that the equivalent strain of the same configuration is only $h_{\rm eq}\simeq\num{6.4e-24}$, far below the reach of any existing wave detector. 
This is however consistent with preliminary estimates of the sensitivity of a rotor to gravitational waves Sec.~\ref{sec:gw_antenna}. 
The coaxial geometry places the detector on the spin axis, where $\widetilde{F}=1$ and the identification with a circularly polarized wave is exact, so that Eq.~\eqref{eq:h_numeric}, which is quoted in the equatorial plane, is reduced here by $\widetilde{F}(\pi/2)=4.65$. The discrepancy with the hour-scale integration time is attributable entirely to the absence of the strain conversion. Above \qty{10}{\mega\hertz} the required integration time becomes prohibitive, because the detector must shrink together with the source and its moment of inertia, which carries the signal quadratically, collapses as $l^{5}$. That regime is better addressed by a detector whose size is decoupled from that of the source, such as a bulk acoustic wave resonator \cite{goryachev2014,campbell2023,omahen_ultracold_2026}.

Note that the two rotors need not in fact be identical. The source enters Eq.~\eqref{eq:N0} only through $Q$ and carries no noise penalty, so it should simply maximize its quadrupole at $\alpha=1/\sqrt{6}$, whereas the detector enters through the ratio $Q^{\rm det\,2}/I_{d}$ and is therefore optimized elsewhere. Maximizing $\alpha^{2}(1-3\alpha^{2})^{2}/(1+3\alpha^{2})$ gives
\begin{equation}
\alpha_{\rm det}^{\ast} = \sqrt{\frac{\sqrt{17}-3}{12}} \simeq 0.306 ,
\label{eq:alpha_det}
\end{equation}
a detector more slender than the source. The quadrupole is reduced by $19\%$ but the moment of inertia, and with it $S_{N}$, by $52\%$, so that $\tau_{1}$ is shortened by a factor $1.36$ at fixed separation. The lengths, by contrast, should be equal, both rotors being pinned at the structural limit of the common rotation frequency. At that common length the new aspect ratio amounts to a detector $25\%$ thinner in radius than the source, and lighter by $44\%$.
While a thinner detector does not deteriorate its structural resistance, being the axial stress of Eq.~\eqref{eq:Omega_max} independent of the radius, it allows for bringing the two bodies closer: the coaxial clearance is now $a+a_{\rm det}$ rather than $2a$, so that contact moves from $\beta=0.816$ to $\beta=0.714$, worth a further factor $3.8$ through the tenth power in Eq.~\eqref{eq:tau1_scaling}.
The entries of Table~\ref{tab:rotor_detector} assume identical rotors and are conservative by this factor. The gain does, however, assume that $\gamma$ is independent of geometry. 

The size of the gain also depends on what dominates the dissipation, since the detector figure of merit is $Q^{\rm det\,2}/\Gamma$, with $S_{N}=4k_{B}T\Gamma$ and $\Gamma$ the drag coefficient. Writing $\Gamma=I_{d}\gamma$ with $\gamma$ independent of geometry, as assumed above, gives Eq.~\eqref{eq:alpha_det}. In the free-molecular regime the drag torque is instead the integral over the rod of a force per unit length growing with the local speed and with the projected width $2a$, so that $\Gamma_{\rm gas}\propto a\,l^{3}$. 
The figure of merit becomes $\alpha^{3}\left(1-3\alpha^{2}\right)^{2}$, the optimum moves to $\alpha_{\rm det}^{\ast}=1/\sqrt{7}\simeq0.378$, and the gain is only $4\%$, the detector being then barely thinner than the source. 
If the loss is instead magnetic, through eddy currents in the surrounding conductors or hysteresis in the levitating superconductor, it scales as the square of the magnetic moment, $\Gamma_{\rm mag}\propto m^{2}\propto a^{4}l^{2}$. This cancels the $a^{4}$ carried by $Q^{\rm det\,2}$ and leaves $\left(1-3\alpha^{2}\right)^{2}$, so the detector prefers to be as slender as possible, the gain reaching $2.1$ already at Eq.~\eqref{eq:alpha_det} and approaching $4$ in the limit. 
That limit is not attainable, since $Q^{\rm det}$ and $\Gamma$ both vanish with $\alpha$ and the aspect ratio is bounded below by whatever noise does not scale with the magnetic moment. In all three cases the detector is thinner than the source, and the transfer of a $\gamma$ measured on one particular rotor to a body of different shape is the delicate step in these estimates.

\subsubsection{Phase coherence and gravitational synchronization}

Coherent integration over a time $\tau$ presupposes that the relative phase in Eq.~\eqref{eq:torque} remains stable over that interval. The same dissipation that sets the torque noise also drives angular momentum diffusion, $\left\langle\Delta L_{z}^{2}\right\rangle = S_{N}\tau/2$, and hence diffusion of the angular position, $\left\langle\Delta\phi^{2}\right\rangle = S_{N}\tau^{3}/6I_{d}^{2}$. The signal is carried at twice that angle, so the quantity that must remain stable is $2\phi$ and the coherence time follows from $\left\langle\Delta(2\phi)^{2}\right\rangle=1$,
\begin{equation}
\left\langle\Delta(2\phi)^{2}\right\rangle = \frac{2S_{N}\tau^{3}}{3I_{d}^{2}}, \qquad \tau_{c} = \left(\frac{3I_{d}^{2}}{2S_{N}}\right)^{1/3} = \left(\frac{3I_{d}}{8k_{B}T\gamma}\right)^{1/3}.
\label{eq:phase_diffusion}
\end{equation}
If both rotors spin freely their diffusion adds and $\tau_{c}$ is shorter by a further $2^{1/3}$, which we neglect on the assumption that the source is constantly driven.

For the \qty{1}{\mega\hertz} entry of Table~\ref{tab:rotor_detector} this gives $\tau_{c}\simeq\qty{3.2e5}{\second}$ at \qty{10}{\milli\kelvin}, some $88$ hours, exceeding $\tau_{1}$ by a factor close to one hundred. The margin is reasonable but it closes rapidly, the two times scaling as $\tau_{c}\propto l^{5/3}$ against $\tau_{1}\propto l^{-5}$. Coherence overtakes the thermal limit at $l\simeq\qty{151}{\micro\metre}$, corresponding to a rotation frequency of \qty{2}{\mega\hertz}. 
At \qty{4.2}{\kelvin} the situation is different, since $\tau_{1}\propto T\gamma$ while $\tau_{c}\propto\left(T\gamma\right)^{-1/3}$. There $\tau_{c}\simeq\qty{4.2e4}{\second}$ at \qty{1}{\mega\hertz}, only three percent of $\tau_{1}$, and the crossing was already at \qty{0.59}{\mega\hertz}, or $l\simeq\qty{505}{\micro\metre}$. Coherent operation at megahertz frequencies therefore requires the lower temperature, and not only because it shortens the integration time.

Beyond the crossing the measurement is no longer coherent over its full duration and must be assembled from segments of length $\tau_{c}$. Each is demodulated coherently to a complex amplitude, and because the relative phase has diffused between segments it is the squared moduli, rather than the amplitudes, that are summed. The minimum detectable torque falls as $\tau^{-1/4}$ rather than $\tau^{-1/2}$, and the time to unit signal-to-noise becomes $\tau\simeq\tau_{1}^{2}/\tau_{c}$.
For the \qty{10}{\mega\hertz} entry at \qty{10}{\milli\kelvin}, where $\tau_{c}\simeq\qty{6.8e3}{\second}$, this amounts to $\sim\num{5e5}$ years. At \qty{1}{\mega\hertz} and \qty{4.2}{\kelvin} the same prescription gives about $\qty{1.4}{\year}$ in place of the $\qty{15.9}{\day}$ of Table~\ref{tab:rotor_detector}, which is the price of operating above the crossing. 

Two mitigation strategies apply near the crossing. The first is an active locking on an external reference frequency. 
The relative phase is independently measurable through the readout of each rotor's angular position, and the measured phase record may be used to correct the demodulation, which extends the usable coherence beyond Eq.~\eqref{eq:phase_diffusion} and is limited in practice by the resolution of the angular readout rather than by thermal diffusion. Actively driven rather than freely spinning rotors are moreover phase locked to their drives, which suppresses the diffusion entirely at the cost of injecting the drive noise in its place. 
A second essentially equivalent scheme consists in a rotational synchronization. By applying an external field oscillating at a reference frequency, a rotor can get effectively synchronized to the external reference, retaining a residual slow librational degree of freedom in the rotating frame. Again, this allows to keep both rotors locked to a given external reference, or shifted with respect to each other by an arbitrary detuning, suppressing relative diffusion at the price of injecting the noise from the reference oscillator. It is worth noting that the synchronization scheme has been effectively demonstrated in the magnetic rotor described in Ref.~\cite{jose2025}. 

A related observation deserves mention. The static torque at $\delta=0$ derives from the interaction energy $U = -U_{0}\cos 2\phi_{0}$ with $U_{0}=N_{0}/2$, so that two freely spinning rotors form a gravitationally bound system, stable at $\phi_{0}=0$ modulo $\pi$, where the two bodies are mutually aligned, the two orientations being physically identical for quadrupolar bodies. 
The configurations $\phi_{0}=\pi/2$ modulo $\pi$ are unstable equilibria. 
Small departures of a single free rotor from the stable orientation librate at
\begin{equation}
\omega_{\rm lib} = \sqrt{\frac{2N_{0}}{I_{d}}} = \sqrt{\frac{\left|\kappa\right| G I_{d}}{2r^{5}}} = \sqrt{\frac{\pi\left|\kappa\right| G\rho}{96\,\beta^{5}}},
\label{eq:libration}
\end{equation}
which in the coaxial configuration at $\beta=2$ gives $\omega_{\rm lib}=\qty{1.8e-5}{\per\second}$, a libration period of about four days. The last form shows that $\omega_{\rm lib}$ depends only on the density and on the separation ratio, so it is the same for every row of Table~\ref{tab:rotor_detector}. 
If both rotors are free the reduced moment of inertia is halved and the frequency is larger by $\sqrt{2}$. Two rotors spun up to nearly equal frequencies and left free would therefore phase lock gravitationally, and the observation of such locking, or of the libration itself, constitutes an alternative signature to the direct torque measurement, with the advantage of being a null-type observation of a frequency rather than an amplitude. 

Two protocols must be distinguished here. 
Spontaneous locking, in which two rotors spun to slightly different rates settle into an aligned state, requires dissipation, since the kinetic energy stored in the frequency mismatch has to be removed for the system to reach a steady state, and without a channel to absorb it the relative orientation keeps advancing at the beat rate. 
The same $\gamma$ that sets the torque noise provides that channel, at the price of a locking time of order $1/\gamma$ (about four months at the value assumed here). 
The alternative is to use synchronization for state preparation alone. Both rotors are first locked to a common external reference, which brings the relative detuning within the trapping condition $\left|\delta\right|\lesssim\sqrt{2}\,\omega_{\rm lib}\simeq\qty{2.6e-5}{\per\second}$, amounting at \qty{1}{\mega\hertz} to a fractional frequency matching of $\num{4e-12}$. 
The pair is then let evolve freely with the phase reference switched off, so that none of its phase noise enters the measurement, which is the principal advantage of this measurement protocol over the active locking one.

What is observed is then the free gravitational libration itself. 
For two identical free rotors the differential angle oscillates at $\sqrt{2}\,\omega_{\rm lib}$ while the common angle is unconstrained, so that the energy of the initial mismatch is exchanged coherently between the two bodies and the rotational line in the readout of each acquires sidebands at $\pm\sqrt{2}\,\omega_{\rm lib}$. 
The requirement for this to be visible is not $\tau_{1}$, which governs the extraction of a torque buried in thermal noise, but that the libration be underdamped, $\gamma\ll\omega_{\rm lib}$, and that the gravitational well be deep compared with the thermal energy, $k_{B}T\ll 2U_{0}=N_{0}$.
At the values assumed here $\omega_{\rm lib}/\gamma\simeq180$, so that some thirty librations elapse before damping and the splitting is resolved by a record a few times longer than the libration period of about three days. 
The first condition is met with a wide margin by the demonstrated hardware, but the second is restrictive. With $N_{0}=\qty{2e-25}{\newton\metre}$ one has $k_{B}T/N_{0}=0.69$ at \qty{10}{\milli\kelvin} and $290$ at \qty{4.2}{\kelvin}, so the relative orientation is only marginally confined in the first case and free in the second. Equivalently, the coherence time Eq.~\eqref{eq:phase_diffusion} is \qty{3.2e5}{\second} at \qty{10}{\milli\kelvin}, longer than the two-rotor libration period $2\pi/\sqrt{2}\,\omega_{\rm lib}=\qty{2.4e5}{\second}$, but only \qty{4.2e4}{\second} at \qty{4.2}{\kelvin}. The synchronization route therefore needs the same millikelvin environment as the direct measurement, and is less demanding only in that it observes a frequency rather than a calibrated torque.
It remains true that at $\delta=0$ the direct measurement is compromised whenever $\omega_{\rm lib}^{-1}=\qty{5.4e4}{\second}$ is not long compared with $\tau_{1}$, which holds at \qty{10}{\milli\kelvin}, where the ratio is $17$, but fails at \qty{4.2}{\kelvin}, where it is four percent, so the linear treatment leading to Eq.~\eqref{eq:SNR} is safe only in the first case. This is a further reason to operate the direct measurement at the small nonzero detuning recommended above, where the phase advances steadily and no equilibrium is approached.

\subsubsection{Systematic couplings}
\label{sec:syscoup}

The thermal limit is not necessarily the dominant obstacle. Any residual electromagnetic coupling between the two rotors exceeds the gravitational torque by many orders of magnitude and thus must be efficiently suppressed.

The most severe case is that of two magnetically levitated rotors, each carrying a permanent dipole $\mathbf{m}$ rotating in the $xy$ plane. For the \qty{1}{\mega\hertz} configuration of Table~\ref{tab:rotor_detector}, a rotor of volume $V=\pi a^{2}l\simeq\qty{1.4e-11}{\cubic\metre}$ with a magnetization of \qty{e6}{\ampere\per\metre}, typical of a rare-earth permanent magnet, carries $|\mathbf{m}|\simeq\qty{1.4e-5}{\ampere\metre\squared}$. The dipole-dipole torque at $r=2l=\qty{600}{\micro\metre}$ is then $\mu_{0}m^{2}/4\pi r^{3}\simeq\qty{9e-8}{\newton\metre}$, exceeding the gravitational torque by nearly eighteen orders of magnitude. Unshielded, the two rotors would librate about mutual magnetic alignment with a period of under a millisecond and would lock for any detuning below $\qty{2e4}{\per\second}$, which is every detuning of interest, so that no beat would survive and the magnetic coupling would dominate.

The measurement is nonetheless not hopeless, because the two couplings appear at different harmonics. 
The dipole-dipole energy is bilinear in the orientation vectors, each of which reverses under a rotation by $\pi$ about the spin axis, so the magnetic torque oscillates at the fundamental detuning $\delta$, whereas the gravitational torque Eq.~\eqref{eq:torque}, bilinear in two quadrupoles that are invariant under the same rotation, oscillates at $2\delta$. 
Reversing the detector by half a turn can thus be used as a rejection strategy, as it changes the sign of the systematic while leaving the signal unchanged.
Since the gravitational torque falls as $r^{-5}$ against $r^{-3}$ for the dipole field, closing the gap to contact improves their ratio by a factor six relative to $\beta=2$.

The separation in frequency is not by itself sufficient, because the magnetic coupling leaks into its own second harmonic. The torque at $\delta$ drives an angular response of the free rotor of amplitude $N_{\rm mag}/I_{d}\delta^{2}$, which on re-entering the coupling generates a component at $2\delta$ of size $N_{\rm mag}^{2}/2I_{d}\delta^{2}$. Requiring this to fall below $N_{0}$ gives
\begin{equation}
N_{\rm mag} \lesssim \delta\,\sqrt{2 I_{d} N_{0}} \simeq \qty{2.2e-20}{\newton\metre}\times\frac{\delta}{\qty{1}{\per\second}} ,
\label{eq:shield_requirement}
\end{equation}
so that some thirteen orders of magnitude of magnetic attenuation are needed at $\delta=\qty{1}{\per\second}$, a detuning that still places a thousand signal cycles within $\tau_{1}$. The requirement relaxes as $\delta^{-1}$, which is a strong argument for operating at the largest detuning the readout bandwidth allows.
Further contamination at $2\delta$ arises from the quadrupolar component of the magnetization distribution and from nonlinearity in the readout, the latter also requiring that the far stronger signal at $\delta$ not saturate it. Superconducting shielding between the two rotors is thus essential, and its attenuation determines the residual crosstalk. The damping it introduces, together with eddy-current and hysteretic loss in the trap, is also what sets $\gamma$ and the effective bath temperature in Eq.~\eqref{eq:SNth}, so shielding cannot be arbitrarily improved.

The alternative of optically or electrically levitated dielectric rotors removes the permanent dipole at the cost of introducing residual charge and patch potentials, which could be addressed by a grounded conducting membrane between the rotors. Such a membrane is transparent to gravity and attenuates electrostatic coupling by many orders of magnitude, its thickness only limited by the requirement that it fits within the gap $r-2a\simeq\qty{355}{\micro\metre}$ left between the two swept envelopes. 
The electrostatic residuals can have different effects. A net charge exerts no torque about a rotor's own centre of mass, and acts on the other body only through that body's charge quadrupole, a coupling that falls as $r^{-3}$ and appears at $2\Omega_{d}$ rather than at $2\delta$. A residual electric dipole produces a similar signature to the magnetic case. The dangerous term being the coupling of the two electric quadrupoles, which sits at $2\delta$ and falls as $r^{-5}$, and is therefore degenerate with the gravitational signal in both respects. Casimir and van der Waals torques scale as steep inverse powers of the separation and are entirely negligible at the hundreds of micrometers considered here, in contrast to short-range gravity experiments at submicron separations.

Residual gas contributes both the damping $\gamma$ entering Eq.~\eqref{eq:SNth} and a direct hydrodynamic coupling between the two rotors. In the free molecular regime the frequency decay rate of a rotor of radius $a$ is
\begin{equation}
\gamma = \frac{10}{\pi}\,\frac{P}{\rho\,\bar{v}\,a} ,
\label{eq:gas_damping}
\end{equation}
following Ref.~\cite{jose2025}, with $\bar v$ the mean thermal speed of the gas. 
Taking $a\simeq l/2$ as an effective radius for the cylinder (Eq.~\eqref{eq:gas_damping} being derived in Ref.~\cite{jose2025} for a sphere) and helium at \qty{4.2}{\kelvin}, for which $\bar v\simeq\qty{149}{\metre\per\second}$, the value $\gamma=\qty{e-7}{\per\second}$ assumed in Table~\ref{tab:rotor_detector} requires a pressure of order \qty{5e-6}{\pascal}. This is an undemanding requirement, since a pressure of \qty{4e-6}{\pascal} has already been reached in Ref.~\cite{jose2025}, where it yielded $\gamma=\qty{4.75e-7}{\per\second}$ on a rotor of radius \qty{24}{\micro\metre}, about six times smaller, for which the same pressure is more detrimental by the factor $1/a$. At that pressure the rotor of Table~\ref{tab:rotor_detector} would be gas limited at $\gamma\simeq\qty{7.7e-8}{\per\second}$. The mean free path is then around \qty{50}{\metre}, exceeding $r$ by nearly five orders of magnitude, so the direct hydrodynamic coupling between the rotors is negligible. 

In summary, three experimental handles discriminate the gravitational signal from the above undesired couplings. The signal follows the second harmonic of the detuning while the dominant electromagnetic couplings follow the first, and stepping or reversing $\delta$ moves the gravitational line in a predicted way. The signal scales as $r^{-5}$ against $r^{-3}$ for dipolar residuals, so that a scan in separation separates the contributions by their exponents. The signal is also proportional to the source quadrupole $Q$, which may be altered while leaving the mass, the magnetic moment and the rotation rate unchanged by exchanging the source rotor for one of different aspect ratio and compensating length. In particular the null case $\alpha=1/\sqrt{3}$ of Eq.~\eqref{eq:Qcyl}, for which $Q$ vanishes identically, reproduces the volume of the \qty{300}{\micro\metre} rotor exactly at $l=\qty{238}{\micro\metre}$, a length still within the structural limit at \qty{1}{\mega\hertz}. Were the length instead left unchanged the volume would double and the dipole-dipole torque quadruple. As noted in Sec.~\ref{sec:cylinder}, such a rotor is an inertially spherical top and therefore has no gyroscopically preferred spin axis, so its orientation must be maintained by the trap alone. This is not an obstacle in the Meissner geometry, where the spin axis is fixed by the anisotropy of the image-dipole interaction and where stable spinning of exactly spherical rotors at megahertz frequencies has been demonstrated \cite{jose2025}. The null test is the only one of the three that also removes the electrostatic quadrupole-quadrupole term, which the other two cannot distinguish from the signal, and it is in our view the most valuable for that reason.


\section{Exploring gravity at megahertz frequencies}
\label{sec:newphys}

Sections~\ref{sec:sources} and \ref{sec:rotor_detector} establish a source and a detector that together generate and measure a near-field gravitational signal at megahertz frequencies. 
This opens the possibility of performing measurements in a frequency range that is currently unexplored, two of which are discussed in detail in this Section. 
The first asks whether the near-field coupling retains its Newtonian form. It is a controlled experiment, employing two rotors, in which the source carries a known quadrupole and a phase, and the observable is a deviation of the measured torque from its predicted amplitude and phase. 
The second asks whether gravitational waves of astrophysical or cosmological origin can be detected at megahertz frequencies, using a single rotor as an antenna. The field is then not controlled, and the sensitivity is accordingly quoted as a noise-equivalent strain rather than as a torque and an integration time.

\subsection{Testing medium-like gravity theories}
\label{sec:mhz_gravity}

The near-field gravitational coupling has been probed up to $\sim$\qty{1.7}{\kilo\hertz} \cite{sinsky1968}, and the measurement proposed here can extend that range by three orders of magnitude. 
What such an experiment would test is the assumption, universal in the interpretation of existing data, that the coupling is instantaneous in the near-field: every experiment performed so far is effectively quasi-static, so no existing measurement could resolve a finite response time of the Newtonian potential.
It is worth clarifying which modified gravity theories this can and cannot discriminate.
For any Lorentz-invariant modification the kernel is a function of $\Box=\eta^{\mu\nu}\partial_\mu\partial_\nu$ alone, so in momentum space the coupling is $\mathcal{G}(k^{2})$ with $k^{2}=-\omega^{2}/c^{2}+|\mathbf{k}|^{2}$. 
In the near-field zone $|\mathbf{k}|\sim1/r\gg\omega/c$, and the frequency enters only through the ratio $\left(\omega r/c\right)^{2}\simeq\num{6e-10}$ for the parameters considered here. The non-local and infinite-derivative completions of general relativity \cite{biswas2012,modesto2012} therefore do predict a frequency-dependent form factor, but one whose frequency dependence at fixed separation is locked to ordinary retardation and is far below any attainable sensitivity. 
A frequency dependence at fixed geometry that is not suppressed by $\left(\omega r/c\right)^{2}$ requires a preferred frame. Scenarios in which gravity is emergent or thermodynamic in origin \cite{jacobson1995,verlinde2011} naturally accommodate one, since the underlying microscopic degrees of freedom constitute a medium, and any medium has a relaxation time on timescales shorter than which the equilibrium description that yields the Einstein equations fails. Those frameworks do not themselves predict its value, which we therefore treat as a free phenomenological parameter. 
The same modification structure arises in dissipative or Lorentz-violating completions, and in any additional gravitational degree of freedom with a dispersive self-interaction. The measurement proposed here is thus best understood as a model-independent null test of medium-like gravity theories.

A parametrization sufficient for reporting a bound is a field relaxation time $\tau_{g}$. 
Interestingly, a heuristic model of retarded gravity with a relaxation time in the millisecond range was proposed as a possible way to resolve the discrepancies between different measurements of $G$ using Cavendish-like experiments \cite{diosi2014}.
For any linear model of relaxation, causality requires the response kernel to vanish for $t<0$ and ties its modulus to its argument through the Kramers-Kronig relations. 
The modification enters the source through a replacement of the Newtonian constant multiplying the time-dependent part of the potential Eq.~\eqref{eq:Phi_quadrupole} by a convolution over the history of the source,
\begin{equation}
G\,Q_{ij}(t) \;\longrightarrow\; \int_{-\infty}^{t}\!dt'\,K(t-t')\,Q_{ij}(t') ,
\label{eq:convolution}
\end{equation}
with $\int_{0}^{\infty}K(t)\,dt=\mathcal{G}_{0}=G$, so that the static monopole term and every quasi-static determination of $G$ are left unchanged. 
The simplest kernel carrying a single timescale is $K(t)=\left(\mathcal{G}_{0}/\tau_{g}\right)e^{-t/\tau_{g}}\Theta(t)$, with $\Theta(t)$ the Heaviside step function, for which a source quadrupole varying as $e^{i\omega t}$ sees the frequency-dependent coupling
\begin{equation}
\mathcal{G}(\omega) = \frac{\mathcal{G}_{0}}{1+i\omega\tau_{g}} ,
\qquad
\left|\mathcal{G}(\omega)\right| = \frac{\mathcal{G}_{0}}{\sqrt{1+\left(\omega\tau_{g}\right)^{2}}} ,
\qquad
\arg \mathcal{G}(\omega) = -\arctan\left(\omega\tau_{g}\right) .
\label{eq:Gomega}
\end{equation}
The field at the detector no longer tracks the instantaneous source distribution but is smeared over a time window $\tau_{g}$, and the non-vanishing imaginary part means that the source does work on the field, as it must be if the underlying degrees of freedom relax. 
Because Eqs.~\eqref{eq:Gij_general}, \eqref{eq:Gijz} and \eqref{eq:torque} are each linear in the source quadrupole, the substitution Eq.~\eqref{eq:convolution} propagates through them unaltered and reaches the observable as $G\to\mathcal{G}(2\Omega)$ in the torque amplitude Eq.~\eqref{eq:N0}. 
Its modulus rescales that amplitude by $\left|\mathcal{G}(2\Omega)\right|/\mathcal{G}_{0}$, while its argument appears as a constant additive offset in the demodulated phase, equivalent to a discrepancy in the relative phase $2\phi_{0}$ of Eq.~\eqref{eq:Qdetector}. 
The two are not independent, being tied to one another by causality, but they are measured independently, so their mutual consistency tests the form Eq.~\eqref{eq:Gomega} itself.

It is worth illustrating the way a torque measurement converts into a bound, since the two channels have different statistics and very different systematics. The signal is a sinusoid at the known frequency $2\delta$ whose amplitude $N_{0}=\kappa\,GQQ^{\rm det}/r^{5}$, Eq.~\eqref{eq:N0}, and phase are both predicted from the geometry and from the independently measured angular positions of the two rotors, and the thermal torque noise $S_{N}$ of Eq.~\eqref{eq:SNth} is white. 
For a sinusoid of known frequency and white noise, the Cram\'er--Rao bounds on amplitude and phase are $\sigma_{N}=\sqrt{2S_{N}/\tau}$ and $\sigma_{\varphi}=\sigma_{N}/N_{0}= \sqrt{\tau_{1} / \tau}$, with $\tau$ the integration time and $\tau_{1}=2S_{N}/N_{0}^{2}$ the time to achieve unit signal-to-noise, Eq.~\eqref{eq:SNR}.
The phase uncertainty in radians is thus equal to the fractional amplitude uncertainty, and both are simply the inverse of the signal-to-noise ratio. 
A modification of gravity enters Eq.~\eqref{eq:torque} as $N_{0}\to\left|\mathcal{G}(2\Omega)\right|N_{0}/\mathcal{G}_{0}$ and as an additive offset $\arg \mathcal{G}(2\Omega)$ in the demodulated phase.
If no deviation is measured within the experimental uncertainty, a bound on $\tau_{g}$ follows from each channel. Expanding Eq.~\eqref{eq:Gomega} for $\omega\tau_{g}\ll1$, the modulus is reduced by $\left(\omega\tau_{g}\right)^{2}/2$ and the argument shifted by $\omega\tau_{g}$, so that requiring each to fall below its own uncertainty gives
\begin{equation}
\tau_{g} \lesssim \frac{1}{\omega}\sqrt{\frac{2\sigma_{N}}{N_{0}}}
\quad\text{(amplitude)},
\qquad
\tau_{g} \lesssim \frac{1}{\omega}\,\frac{\sigma_{N}}{N_{0}}
\quad\text{(phase)}.
\label{eq:taug_bound}
\end{equation}
The phase is more sensitive by a factor $\sqrt{2N_{0}/\sigma_{N}}$, simply because it responds at first order in $\omega\tau_{g}$ while the amplitude responds only at second order. Both bounds carry the same factor $1/\omega$, and since $N_{0}$ and $S_{N}$ are each independent of the rotation rate at fixed geometry, so is $\sigma_{N}/N_{0}$ at fixed integration time.
Raising the rotation frequency by a factor of ten therefore improves the bound on $\tau_{g}$ by the same factor at no cost in measurement time, up to the structural limit of the rotor in use, beyond which a smaller rotor is required and the integration time grows as $\Omega^{5}$ by Eq.~\eqref{eq:tau1_scaling}.

Both bounds involve measurement results only through $\sigma_{N}/N_{0}=\mathrm{SNR}^{-1}=\sqrt{\tau_{1}/\tau}$, so that Eq.~\eqref{eq:taug_bound} may be written entirely in terms of the integration time in units of the $\tau_{1}$ of Table~\ref{tab:rotor_detector},
\begin{equation}
\tau_{g} \lesssim \frac{\sqrt{2}}{\omega}\left(\frac{\tau_{1}}{\tau}\right)^{1/4}
\quad\text{(amplitude)},
\qquad
\tau_{g} \lesssim \frac{1}{\omega}\left(\frac{\tau_{1}}{\tau}\right)^{1/2}
\quad\text{(phase)} .
\label{eq:taug_tau}
\end{equation}
Evaluating at the \qty{1}{\mega\hertz} row of Table~\ref{tab:rotor_detector}, that is at a signal frequency $\omega/2\pi=\qty{2}{\mega\hertz}$, and inverting Eq.~\eqref{eq:Gomega} exactly rather than using Eq.~\eqref{eq:taug_bound}, which is necessary once $\omega\tau_{g}$ approaches unity, a measurement at $\mathrm{SNR}=10$ gives $\tau_{g}\lesssim\qty{3.9e-8}{\second}$ from the amplitude and $\qty{8.0e-9}{\second}$ from the phase. 
This requires $\tau=100\,\tau_{1}$, which is \qty{3.8}{\day} at \qty{10}{\milli\kelvin} and \qty{4.4}{\year} at \qty{4.2}{\kelvin}. 
For reference, the light-crossing time of the apparatus is $r/c=\qty{2}{\pico\second}$ at $r=\qty{600}{\micro\metre}$, below which any such measurement would be limited by ordinary retardation rather than by new physics. 
The only existing constraint of this kind comes from Ref.~\cite{sinsky1968}, where agreement with Newtonian theory at \qty{1660}{\hertz}, taken here at the \qty{10}{\percent} level in amplitude, corresponds to $\tau_{g}\lesssim\qty{5e-5}{\second}$. 
That experiment had no phase reference, so the comparison on equal terms is with the amplitude channel, on which the present scheme improves by a factor $\num{1.2e3}$ at $\mathrm{SNR}=10$.
The phase channel available here would extend the improvement to $\num{5.8e3}$ at the same signal-to-noise level.

Several features make the phase the more informative observable. 
The first is intrinsic sensitivity, since by Eq.~\eqref{eq:Gomega} the argument departs from zero linearly in $\omega\tau_{g}$ whereas the modulus departs quadratically, so for a small deviation the phase channel is stronger by $\sqrt{2N_{0}/\sigma_{N}}$ regardless of how well the apparatus is calibrated. 
The second is that the phase is obtained from the relative angular positions of the two rotors, which can be measured to high accuracy by the transducers, rather than to an absolute torque calibration. 
The amplitude channel by contrast requires $QQ^{\rm det}/r^{5}$ to be known to the same fractional accuracy as the measurement itself, which at $\mathrm{SNR}=10$ demands $\delta r/r\lesssim\num{2e-2}$ from the separation alone, or \qty{12}{\micro\metre} on \qty{600}{\micro\metre}.
The phase offset is, however, degenerate with the encoder zero $\phi_{0}$ and with the fixed phase response of the readout chain, so it cannot be interpreted absolutely in a single measurement. The degeneracy is broken by holding the detuning $\delta$ fixed and sweeping the common rotation frequency. The term $\arctan(2\Omega\tau_{g})$ grows with $\Omega$, whereas an encoder offset does not, and since the demodulated signal stays at the same frequency $2\delta$ throughout, the readout phase is constant along the scan.

The same sweep makes the scheme frequency-tunable in a way that resonant experiments are not. The rotation frequency of a levitated rotor can be continuously varied by synchronous driving, from the trap libration frequency up to the structural disintegration limit \cite{jose2025}. A single apparatus can therefore map the gravitational coupling continuously from kilohertz to megahertz, with the predicted signal amplitude and the integration time to a given signal-to-noise both constant along the scan at fixed geometry, so that any frequency dependence appears directly as a departure from a flat response in amplitude and as a growing offset in phase.

\subsection{Detecting high-frequency gravitational waves}
\label{sec:gw_antenna}

The single-rotor detector of Sec.~\ref{sec:rotor_detector} responds to any tidal field oscillating near twice its own spin rate, which may be either the near field of a second rotor or the far field of a passing gravitational wave. 
In the latter case the source is no longer laboratory-based but of astrophysical or cosmological origin. 
For this application, the figure of merit is not a torque but rather the noise-equivalent strain of Ref.~\cite{aggarwal2025}.

The response follows from Sec.~\ref{sec:heq}.
A circularly polarized wave of amplitude $h$ and angular frequency $\omega$ propagating along the spin axis presents the tidal tensor $G^{\rm GW}_{ij}=\tfrac{1}{2}\ddot{h}_{ij}$, of amplitude $\omega^{2}h/2$ and of exactly the rotating form of Eq.~\eqref{eq:Gijz}.
Inserting it into Eq.~\eqref{eq:torqueG} with the detector quadrupole Eq.~\eqref{eq:Qdetector} reproduces the coaxial case of Sec.~\ref{sec:rotor_detector} with $\kappa_{-}=2/3$, the torque appearing at the difference frequency $\left|\omega-2\Omega_{d}\right|$ with amplitude
\begin{equation}
N^{\rm GW}_{0} = \frac{1}{3}\,\omega^{2}\,h\,Q^{\rm det} = \frac{\omega^{2}I_{d}}{6}\,h ,
\label{eq:N_GW}
\end{equation}
where the second equality uses $Q^{\rm det}=I_{d}/2$, valid for the optimal cylinder of Sec.~\ref{sec:cylinder}.
The rotor is thus operated at half the frequency of the wave it detects, $\Omega_{d}\simeq\omega/2$, exactly as the source rotor emits at twice its spinning frequency.
Dividing the thermal torque noise Eq.~\eqref{eq:SNth} by the square of the responsivity $N^{\rm GW}_{0}/h$ of Eq.~\eqref{eq:N_GW} gives the single-sided noise-equivalent strain spectral density
\begin{equation}
S^{\rm noise}_{h}(\omega) = \frac{144\,k_{B}T\gamma}{\omega^{4}I_{d}} \;.
\label{eq:Sh_noise}
\end{equation}
The $\omega^{-4}$ is the mirror image of the $\omega^{-2}$ of Sec.~\ref{sec:heq}. 
A wave of fixed strain presents a tidal gradient growing as $\omega^{2}h$, so the responsivity Eq.~\eqref{eq:N_GW} grows quadratically with frequency at fixed rotor.
It does not follow that the antenna improves with frequency, since the rotor must spin at $\Omega_{d}=\omega/2$ and the structural limit then ties its inertia to the frequency as well.
The $I_{d}^{-1}$, in turn, is weaker than the quadratic dependence on $Q^{\rm det}$ inherited from the square of Eq.~\eqref{eq:N_GW} would suggest, because at fixed $\gamma$ the dissipation grows with the inertia in the same proportion, $S_{N}\propto I_{d}$ by Eq.~\eqref{eq:SNth}, and only one power survives in the ratio.
What it does imply is that the rotor should be the largest one that can still be spun at $\omega/2$, which is why Eq.~\eqref{eq:Sh_noise} is to be evaluated at the structural limit.

Writing Eq.~\eqref{eq:Omega_max} at $\Omega_{\max}=\omega/2$ gives $l=4v_{\rm tip}/\omega$ with $v_{\rm tip}=\sqrt{2\sigma_{\rm break}/\rho}$ the tip speed at failure, and eliminating the size in favor of the frequency in Eq.~\eqref{eq:Sh_noise} leaves
\begin{equation}
S^{\rm noise}_{h}(\omega) = \frac{27}{4\pi}\,\frac{k_{B}T\gamma\,\omega}{\rho\,v_{\rm tip}^{5}} = \frac{27}{2^{9/2}\pi}\,k_{B}T\gamma\,\omega\,\frac{\rho^{3/2}}{\sigma_{\rm break}^{5/2}} ,
\label{eq:Sh_limit}
\end{equation}
so that $\sqrt{S^{\rm noise}_{h}}$ grows only as $\omega^{1/2}$.
The sensitivity is therefore nearly flat across the band, the $\omega^{-5}$ of the shrinking inertia being largely compensated by the $\omega^{4}$ of Eq.~\eqref{eq:Sh_noise}.
The flatness is however inherited from the assumption $\Gamma=I_{d}\gamma$ made in Eq.~\eqref{eq:SNth}.
Writing instead $S^{\rm noise}_{h}\propto\Gamma/\omega^{4}Q^{\rm det\,2}$ with the drag coefficients of Sec.~\ref{sec:rotor_detector}, free-molecular gas drag $\Gamma_{\rm gas}\propto a\,l^{3}$ gives $\sqrt{S^{\rm noise}_{h}}\propto\omega$, whereas magnetic loss $\Gamma_{\rm mag}\propto a^{4}l^{2}$ cancels the frequency dependence altogether and leaves the noise-equivalent strain exactly flat.
The band over which the sensitivity is nearly constant thus depends on both the geometry and on the dominant dissipation channel.

The material dependence must be evaluated at fixed $\omega$ with the rotor at the structural limit, where $l=4v_{\rm tip}/\omega$, and not through the figure of merit $\rho^{2}/\sigma_{\rm break}$ of Eq.~\eqref{eq:h_material}, which is written at fixed length and therefore compares two materials at two different frequencies. 
The inertia available at a given frequency is $I_{d}\propto\rho\,l^{5}\propto\sigma_{\rm break}^{5/2}\rho^{-3/2}$, and since the antenna responds linearly in it, Eq.~\eqref{eq:Sh_limit} improves as $S_{h}^{\rm noise\,-1}\propto\sigma_{\rm break}^{5/2}\rho^{-3/2}$. This implies that both a strong and a light material help, the latter because it permits a longer rotor at the same spin rate. 
Such density preference is opposite to the one for the two-rotor experiment of Sec.~\ref{sec:rotor_detector}, where at fixed $r/l$ one has $\tau_{1}^{-1}\propto\rho^{3}l^{5}\propto\sigma_{\rm break}^{5/2}\rho^{1/2}$, the four powers of density carried by the squared quadrupole product outweighing the single power in the inertia.

Equation~\eqref{eq:Sh_limit} depends on the breaking stress as $\sigma_{\rm break}^{-5/4}$ in amplitude, so the uncertainty discussed in Sec.~\ref{sec:limits} propagates directly and we quote both values.
Table~\ref{tab:gw_antenna} evaluates Eq.~\eqref{eq:Sh_limit} for NdFeB at the $T=\qty{10}{\milli\kelvin}$ and $\gamma=\qty{e-7}{\per\second}$ of Table~\ref{tab:rotor_detector}, on the calibrated estimate $\sigma_{\rm break}=\qty{3.4}{\giga\pascal}$ of Refs.~\cite{schuck2018,jose2025} and on the \qty{556}{\mega\pascal} actually sustained in Ref.~\cite{jose2025}.
The two stresses differ by a factor $6.1$ and the resulting noise-equivalent strains by $9.6$.
Coherent integration converts these into a minimum detectable strain $h_{\min}=\sqrt{2S^{\rm noise}_{h}/\tau}$, which at \qty{2}{\mega\hertz} and over the coherence time $\tau_{c}$ of Eq.~\eqref{eq:phase_diffusion} gives $h_{\min}\simeq\num{6.2e-25}$ on the optimistic material, for which $\tau_{c}\simeq\qty{3.2e5}{\second}$ as in Sec.~\ref{sec:rotor_detector}, and $\num{1.3e-23}$ on the demonstrated one, where the shorter rotor carries $\tau_{c}$ down to $\qty{7.2e4}{\second}$ through $\tau_{c}\propto l^{5/3}$.
The penalty between the two scenarios is therefore a factor $20$ rather than the $9.6$ of the spectral density alone, the breaking stress entering both the sensitivity and the time over which it may be exploited.

\renewcommand{\arraystretch}{1.8}
\setlength{\tabcolsep}{10pt}

\begin{table}[t]
\centering
\begin{tabular}{ccccc}
\hline\hline
 & \multicolumn{2}{c}{$\sigma_{\rm break}=\qty{3.4}{\giga\pascal}$} & \multicolumn{2}{c}{$\sigma_{\rm break}=\qty{556}{\mega\pascal}$} \\
$f=\omega/2\pi$ & $l$ & $\sqrt{S^{\rm noise}_{h}}$ & $l$ & $\sqrt{S^{\rm noise}_{h}}$ \\
\hline
\qty{0.2}{\mega\hertz} & \qty{3.00}{\milli\metre}  & \qty{7.9e-23}{\rtHz} & \qty{1.23}{\milli\metre}  & \qty{7.6e-22}{\rtHz} \\
\qty{2}{\mega\hertz}   & \qty{300}{\micro\metre}   & \qty{2.5e-22}{\rtHz} & \qty{123}{\micro\metre}   & \qty{2.4e-21}{\rtHz} \\
\qty{20}{\mega\hertz}  & \qty{30.0}{\micro\metre}  & \qty{7.9e-22}{\rtHz} & \qty{12.3}{\micro\metre}  & \qty{7.6e-21}{\rtHz} \\
\hline\hline
\end{tabular}
\caption{\textbf{Noise-equivalent strain of a single levitated NdFeB rotor operated at its structural limit}, from Eq.~\eqref{eq:Sh_limit} at $T=\qty{10}{\milli\kelvin}$ and $\gamma=\qty{e-7}{\per\second}$, for the two breaking stresses discussed in Sec.~\ref{sec:limits}. The rotor length is $l=4v_{\rm tip}/\omega$, rounded as in Table~\ref{tab:rotor_detector}, and the rotation frequency is $f/2$. The two blocks differ by the fixed factor $9.6=(3400/556)^{5/4}$ and may be rescaled to any other temperature and damping rate through $\sqrt{S^{\rm noise}_{h}}\propto\left(T\gamma\right)^{1/2}$.}
\label{tab:gw_antenna}
\end{table}

For comparison, bulk acoustic wave detectors are estimated at $\sqrt{S_{h}}\simeq\qty{e-22}{\rtHz}$ per mode \cite{goryachev2014}, the MAGE experiment operates with a spectral sensitivity as low as \qty{6.6e-21}{\rtHz} in narrow bands across the megahertz range \cite{campbell2023}, and the levitated-sensor detector targets $\qty{e-21}{\rtHz}$ at \qty{100}{\kilo\hertz}, with $\qty{e-22}{\rtHz}$ projected for a ten-metre instrument \cite{aggarwal2022}.
A single levitated microrotor is thus competitive with the best projections in this band.
Since $\sqrt{S^{\rm noise}_{h}}\propto\rho^{3/4}\sigma_{\rm break}^{-5/4}$, silica would gain a further factor $4$ by being at once stronger and lighter, the cost being optical rather than magnetic levitation, for which the ultra-low damping rate assumed here is not established.

The bandwidth of a rotor detector is of a different nature from that of a resonant antenna, as we now discuss.
The responsivity Eq.~\eqref{eq:N_GW} carries no mechanical resonance, the detector acting as a heterodyne receiver whose local oscillator is the rotation itself.
A freely spinning rotor responds to a torque at offset $\Delta\omega$ from $2\Omega_{d}$ with an angular response $\Delta \theta =N/I_{d}\Delta\omega^{2}$, and since the thermal torque noise is converted by the same factor, the ratio of signal to thermal noise is independent of $\Delta\omega$ and Eq.~\eqref{eq:Sh_limit} holds at any offset.
What limits the bandwidth is the angular readout. Equating the thermal angle noise to a readout angle noise $S_{\theta}$, taken white, gives
\begin{equation}
\Delta\omega_{\max} = \left(\frac{S_{N}}{I_{d}^{2}S_{\theta}}\right)^{1/4} = \left(\frac{4k_{B}T\gamma}{I_{d}S_{\theta}}\right)^{1/4} ,
\label{eq:gw_band}
\end{equation}
which is a free-mass limit rather than a resonance limit.
For the \qty{2}{\mega\hertz}, \qty{3.4}{\giga\pascal} entry of Table~\ref{tab:gw_antenna} and a reference angular readout noise $\sqrt{S_{\theta}}=\qty{e-9}{\radian\rtHz}$, Eq.~\eqref{eq:gw_band} gives $\Delta\omega_{\max}\simeq\qty{2.6}{\per\second}$, that is, a full band $\Delta\omega_{\max}/\pi$ of order one hertz about the carrier, which is how we quote bands below.
This is narrow in absolute terms but not in relative ones. In fact, the fractional band $2\Delta\omega_{\max}/\omega\simeq\num{4e-7}$ exceeds the $1/Q_{m}\simeq\num{e-9}$ of a bulk acoustic resonator by more than two orders of magnitude.
Since $I_{d}\propto\omega^{-5}$ at the structural limit, the band widens as $\Delta\omega_{\max}\propto\omega^{5/4}$, reaching a band of some \qty{15}{\hertz} at \qty{20}{\mega\hertz} and falling to a few tens of millihertz at \qty{0.2}{\mega\hertz}.
It also widens as $\left(T\gamma\right)^{1/4}$, so that the cooling and the loss reduction which improve Eq.~\eqref{eq:Sh_limit} narrow the band, and the two must be traded against one another.
The centre frequency is in any case tunable from the trap libration frequency up to the structural limit, so that a single device can scan a wide band rather than sampling it at fixed overtones.

Three caveats attach to our estimates.
Equation~\eqref{eq:Sh_limit} is written for a circularly polarized wave incident along the spin axis, for which the identification of Sec.~\ref{sec:heq} is exact. 
Incidentally, we note that many astrophysical sources, including the binaries so far detected at low frequencies, do radiate circularly polarized waves along their orbital axis, and that a rotor is in this respect unusual in being naturally matched to a circular polarization.
For linearly polarized waves, only the co-rotating circular component downconverts to the readout band, the counter-rotating one appearing at $\omega+2\Omega_{d}$, so a linearly polarized wave of amplitude $h$ drives the torque at half that amplitude and degrades $\sqrt{S^{\rm noise}_{h}}$ of Eq.~\eqref{eq:Sh_limit} by a factor two. 
A general direction of incidence introduces in addition an antenna pattern of order unity which we do not evaluate here.
The coherence analysis of Sec.~\ref{sec:rotor_detector} applies unchanged. The recorded angular position removes the rotor's own phase diffusion of Eq.~\eqref{eq:phase_diffusion}, leaving the coherence of the wave itself as the limit, which for an unmodelled astrophysical signal is not known in advance.
Finally, the low-frequency entry of Table~\ref{tab:gw_antenna} requires a rotor of millimetre size, whose stable levitation presents the same practical obstacle that sets the lower end of the useful range in Sec.~\ref{sec:rotor_detector}, so the \qty{0.2}{\mega\hertz} row should be read as an extrapolation rather than as a configuration in hand.

It is worth briefly discussing what type of source a rotor-based detector at megahertz frequencies could target. An exhaustive discussion of potential sources of gravitational waves at megahertz frequencies and above is given in Ref.~\cite{aggarwal2022}.
A rotor detector features a relatively narrow band of a few hertz around twice the spinning frequency, limited by the angular detection noise. In this respect, it would behave as other narrowband detectors, such as bulk acoustic resonators \cite{goryachev2014,omahen_ultracold_2026} or optically levitated detectors \cite{aggarwal2022}. In particular, it would be similarly sensitive to impulsive broadband gravitational waves, such as those arising from mergers of primordial black holes. 
However, narrowband detectors are far from optimal for broadband sources, since they are sensitive only to the fraction of the broadband signal falling in the narrow sensitivity band.

The high tunability and compactness suggest that a rotor detector could be more suitable to the search for a stochastic background of gravitational waves, which can be of primordial origin or arise from exotic mechanisms \cite{aggarwal2025}. Detection of a stochastic background requires coherent cross-correlation of two or more detectors within the same sensitivity band, preferably with the same orientation and placed at a distance much shorter than the wavelength, which for $f=\qty{2}{\mega\hertz}$ corresponds to $\sim\qty{150}{\metre}$. The compactness and tunability of a rotor would allow $n\gg1$ detectors to be operated in the same laboratory, with the same orientation and phase-locked to the same external reference. The stability of this configuration allows in principle for an indefinite integration time $T$. The minimum detectable noise power spectral density of a stochastic background then scales as \cite{allen1999}:
\begin{equation}
  S_{h,\mathrm{st}}\simeq \frac{S_h}{\sqrt{n_p T\Delta f}} ,
\label{eq:stochasticGW}
\end{equation}
where $n_p$ stands for the number of independent detector pairs, $n_p=n(n-1)/2$ for $n$ detectors.
For instance, taking the optimistic central line in Table~\ref{tab:gw_antenna} with signal at \qty{2}{\mega\hertz}, with $n=20$, $\Delta f=\qty{1}{\hertz}$ as given by Eq.~\eqref{eq:gw_band} and $T=\qty{1}{\year}$, the minimum detectable strain spectral density from a stochastic background would be as low as $S_{h,\mathrm{st}}^{1/2}\simeq\qty{9e-25}{\rtHz}$, well beyond the reach of other proposed detectors in this band.

\section{Conclusions}

We have shown that a pair of levitated microrotors can generate and measure a gravitational interaction at megahertz frequencies, three orders of magnitude above previously reported experiments. The essential ingredient come from the fact that a detector rotor can perform a frequency conversion of the signal mechanically. By using its rotation as a local oscillator, a field modulated at several megahertz can be read out as a torque at a few hertz. 
In addition, for a fixed geometry it turns out that neither the predicted torque nor the time needed to resolve the signal depends on how fast the rotors turn, so that a gravitational signal can be generated at whatever frequency the material allows, up to the disintegration limit, and still be read out in the same band of a few hertz at no cost in sensitivity. For the case of two coaxial NdFeB cylinders of \qty{300}{\micro\metre} separated by twice their length, each spinning at \qty{1}{\mega\hertz} so that the signal appears at \qty{2}{\mega\hertz}, the torque is \qty{2e-25}{\newton\metre} and a unit signal-to-noise is reached in \qty{55}{\minute} at \qty{10}{\milli\kelvin}. This timescale improves as the fifth power of the rotor size and would shorten by three orders of magnitude at half the separation. Interpreted as a bound on a single relaxation time of the gravitational field, a measurement at $\mathrm{SNR}=10$ would reach the \qty{e-8}{\second} level, with the phase of the demodulated torque providing the most sensitive observable as it responds already at first order in $\omega\tau_{g}$. The same apparatus, operated with a single rotor as a gravitational wave antenna, has a noise-equivalent strain of a few $10^{-22}\,\mathrm{Hz}^{-1/2}$ in a band of about a hertz that can be tuned anywhere between the trap libration frequency and the disintegration limit.

As a roadmap for the proposed experiments, a natural first step would be to test a few of the assumptions on which our estimated numbers rely. From the geometrical perspective, since the integration time varies as the tenth power of the separation in units of the rotor length, more refined estimates would require exact finite-body calculations of the gravitational interaction at small separations. Next comes the breaking stress of NdFeB at the hundred-micrometre scale, which enters the torque as $\sigma_{\rm break}^{5/2}$ and which, rather than being inferred, could be measured directly by spinning a rotor to destruction. The electromagnetic environment remains the hardest practical obstacle. The two couplings do live at different harmonics of the relative phase and respond differently to a half-turn reversal of one rotor, but second-order leakage into the signal band still leaves a demanding superconducting shielding requirement. Finally, the millikelvin operation presupposes a dissipative environment at that temperature for a body that is thermally isolated by construction and that carries a large rotational energy, so that passive thermalization may not suffice and active cooling of the librational degree of freedom could be required.

In conclusion, we believe that the proposed experiment could give observational access to gravitational phenomenology in a regime so far unexplored. The examples we have analyzed include a direct bound on a relaxation time of the near-field coupling in the megahertz range, which would constrain the emergent, thermodynamic and dissipative completions in which gravity inherits a response time from an underlying medium. Operated as a single-rotor antenna, the same device becomes a tunable receiver in a band where no astrophysical source is guaranteed, but where a detection could point to exotic physics.

\vspace{3mm}
\textbf{Acknowledgments. --} MF was supported by the Swiss National Science Foundation Ambizione Grant No. 208886, and by The Branco Weiss Fellowship -- Society in Science, administered by the ETH Z\"{u}rich. 
The authors would like to acknowledge networking support by the COST Action CA23130 ``BridgeQG: Bridging high and low energies in search of quantum gravity''.

\clearpage
\newpage
\bibliography{references}

\end{document}